\documentclass[sigconf]{acmart}

\AtBeginDocument{%
  }

\copyrightyear{\textbf{2024}}
\acmYear{2024}
\setcopyright{acmlicensed}
\acmConference[UIST '24]{The 37th Annual ACM Symposium on User Interface Software and Technology}{October 13--16, 2024}{Pittsburgh, PA, USA}
\acmBooktitle{The 37th Annual ACM Symposium on User Interface Software and Technology (UIST '24), October 13--16, 2024, Pittsburgh, PA, USA}
\acmDOI{10.1145/3654777.3676464}
\acmISBN{979-8-4007-0628-8/24/10}

\usepackage{dblfloatfix}
\usepackage{enumitem}

\begin{document}

\title{Degrade to Function: Towards Eco-friendly Morphing Devices that Function Through Programmed Sequential Degradation}

\author{Qiuyu Lu}
\orcid{0000-0002-8499-3091}
\authornote{Both authors contributed equally to this research.}
\authornote{qiuyulu@berkeley.edu, seminay@andrew.cmu.edu, liningy@berkeley.edu}
\affiliation{%
  \institution{University of California, Berkeley}
  \streetaddress{}
  \city{Berkeley}
  \state{CA}
  \country{USA}
  \postcode{94720}
  }


\author{Semina Yi}
\authornotemark[1]
\authornotemark[2]
\affiliation{%
  \institution{Carnegie Mellon University}
  \streetaddress{5000 Forbes Ave}
  \city{Pittsburgh}
  \state{PA}
  \country{USA}}

\author{Mengtian Gan}
\affiliation{%
  \institution{Carnegie Mellon University}
  \streetaddress{5000 Forbes Ave}
  \city{Pittsburgh}
  \state{PA}
  \country{USA}}

\author{Jihong Huang}
\affiliation{%
  \institution{Carnegie Mellon University}
  \streetaddress{5000 Forbes Ave}
  \city{Pittsburgh}
  \state{PA}
  \country{USA}}

\author{Xiao Zhang}
\affiliation{%
  \institution{Tsinghua University}
  \city{Beijing}
  \country{China}}

\author{Yue Yang}
\affiliation{%
  \institution{University of California, Berkeley}
  \city{Berkeley}
  \state{CA}
  \country{USA}}

\author{Chenyi Shen}
\affiliation{%
  \institution{Carnegie Mellon University}
  \city{Pittsburgh}
  \state{PA}
  \country{USA}}

\author{Lining Yao}
\authornotemark[2]
\affiliation{%
  \institution{University of California, Berkeley}
  \city{Berkeley}
  \state{CA}
  \country{USA}}

\renewcommand{\shortauthors}{Lu, et al.}
\renewcommand{\shorttitle}{Degrade to Function}

\begin{abstract}
While it seems counterintuitive to think of degradation within an operating device as beneficial, one may argue that when rationally designed, the controlled breakdown of materials— physical, chemical, or biological—can be harnessed for specific functions. To apply this principle to the design of morphing \textcolor{black}{devices}, we introduce the concept of "Degrade to \textcolor{black}{Function}" (\textcolor{black}{DtF}). This concept aims to create eco-friendly and self-contained morphing \textcolor{black}{devices} that operate through a sequence of environmentally-triggered degradations. We explore its design considerations and implementation techniques by identifying environmental conditions and degradation types that can be exploited, evaluating potential materials capable of controlled degradation, suggesting designs for structures that can leverage degradation to achieve various transformations and functions, and developing sequential control approaches that integrate degradation triggers. To demonstrate the viability and versatility of this design strategy, we showcase several application examples across a range of environment conditions.
\end{abstract}



\begin{teaserfigure}
  \includegraphics[width=\textwidth]{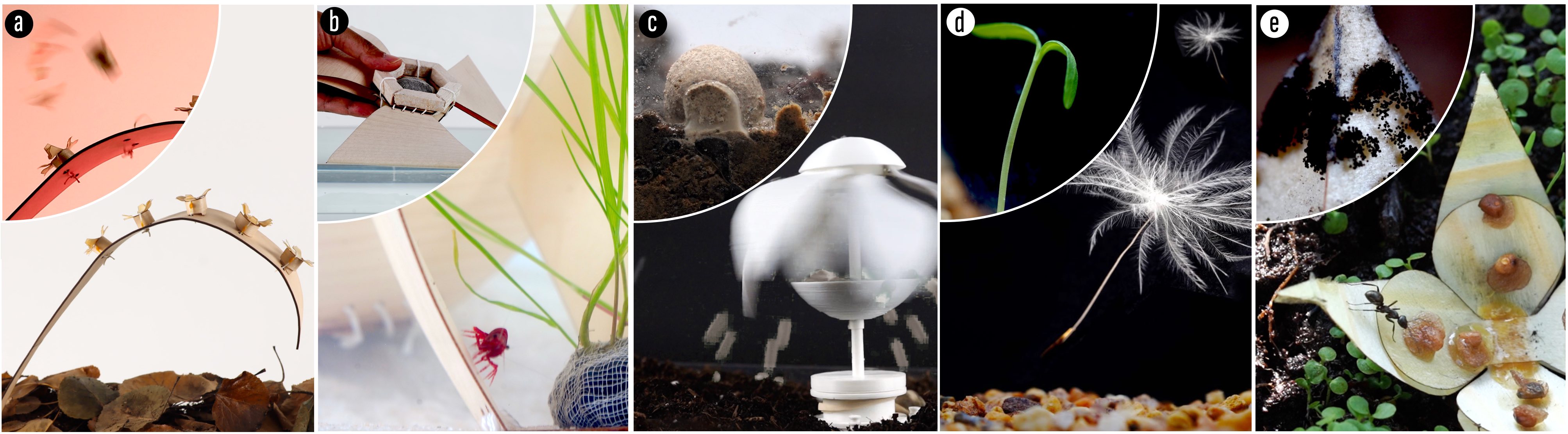}
  \caption{Presenting “Degrade to Function” examples: the self-sustaining wildfire monitor (a), aquatic shelter (b), acidic soil restorer (c), desertification mediator (d), and forest seeder (e). Each is capable of functional sequential transformations, activated by accelerated partial structure degradation in response to specific environmental conditions.}
  \Description{}
  \label{fig:teaser}
\end{teaserfigure}


\maketitle

\section{Introduction}
Sustainable design has emerged as an significant theme within the field of Human-Computer Interaction (HCI), in material exploration, interface design and device fabrication. Research \textcolor{black}{has} dedicated considerable efforts towards the development of easy-to-unmake techniques. Strategies such as recycling, which involves the repurposing of materials that are no longer in use \cite{wall_scrappy_2021, bell_reclaym_2022}, and upcycling, characterized by the creative reutilization of materials \cite{hanton_fabricatink_2022, williams_upcycled_2021}, have become central to this endeavor. Equally important is the facilitation of easy disassembly to support these processes \cite{song_unmaking_2021, wu_unfabricate_2020, murer_-crafting_2017}. While these strategies have contributed to mitigating resource-related challenges, they alone cannot fully resolve waste-related issues. Consequently, there is a growing interest in leveraging the materials' (bio)degradability at their end-of-life stage, spotlighting it as a potent avenue for sustainable waste management. This includes investigating the utilization of food-grade organic materials, such as flour \cite{buechley_3d_2023}, coffee grounds \cite{rivera_designing_2023}, and compost \cite{bell_reclaym_2022}, naturally sourced materials like leaves \cite{song_towards_2022} and beeswax \cite{cheng_functional_2023}, as well as biodegradable polymers such as PVA \cite{arroyos_tale_2022}, PHA \cite{meereboer_review_2020}, and PLA \cite{sangkasanya_feasibility_2018} in prototyping and packaging, reconfiguring and repurposing. 

\textcolor{black}{Degradation generally signifies the gradual breakdown or deterioration of materials, which reduces the quality, value, or functionality over time \cite{Vásquez-Grandón_Forest_2018}}. While degradation is commonly viewed as detrimental to device operation and ideally only occurs at the end-of-life \cite{hartmann_becoming_2021}, there are numerous cases where the physical, chemical, or biological breakdown of materials is intentionally leveraged for a specific function. For instance, \textcolor{black}{fuses safeguard against electrical overloads \cite{EDISON_1890}, and sacrificial anodes are used to prevent metal corrosion \cite{pedeferri_cathodic_1996}. Similarly purposeful degradation is evident in nature: the squirting cucumber's seed capsules burst to release seeds \cite{galstyan_snap_2018}, and black wattle seeds crack only under fire heat, aiding in post-fire germination \cite{tangney_success_2022}}. Adopting a similar strategy, Functional Destruction \cite{cheng_functional_2023} suggests that material destruction can be strategically used to fulfill specific objectives of the transient electronics.


On the other hand, although electronics can incorporate degradable materials, they usually still depend on conventional components such as semiconductors, microcontrollers, and batteries for functionality \cite{koelle_prototyping_2022, vasquez_myco-accessories_2019}. Yet, the realm of morphing \textcolor{black}{devices} has seen the rise of electronic-free devices as a complement. These devices often utilize innovative materials that can respond to various stimuli to create functional mechanisms, including shape-memory material for heat-responsive \textcolor{black}{devices} \cite{zhong_epomemory_2023, wang_printed_2018, wang_-line_2019}, synthetic hydrogels for pH-responsive mini-robots \cite{xin_environmentally_2021} or display \cite{kan_organic_2017}, and bacteria-coated films \cite{yao_biologic_2015} and chemically-treated wood veneers \cite{luo_autonomous_2023} for moisture-triggered devices \cite{feng_cholesteric_2023}, etc. Nonetheless, these advanced materials can be expensive and complicated to manufacture. They may also require specialized recycling processes, resulting in high embodied energy and carbon dioxide emissions throughout their lifecycle \cite{koelle_prototyping_2022}. This could compromise efforts to achieve sustainability goals.

Inspired by these precedents and aiming to address existing gaps, we explore the concept of "Degrade to \textcolor{black}{Function}" (\textcolor{black}{DtF}). \textcolor{black}{This approach invites HCI researchers and designers to explore the potential of utilizing the end-of-life phase of materials as an active period for morphing \textcolor{black}{devices}}.
We investigate natural materials that exhibit different rates of degradation under varying environmental conditions. By strategically integrating these degradation characteristics with environmental condition changes, we aim to design autonomous morphing \textcolor{black}{devices} capable of sequential deformations to achieve specific functions. We hope this research spurs the development of sustainable morphing \textcolor{black}{devices} that are not only environmentally friendly but also versatile in their applications, particularly in ecology-related scenarios. The core contributions of this work are as follows:

\begin{itemize}[leftmargin=10pt]
    \item Introducing the \textcolor{black}{DtF} design strategy, aimed at creating sustainable, self-contained morphing \textcolor{black}{devices} that operate through environmentally triggered sequential degradation.  
    \item Exploring the \textcolor{black}{DtF} design considerations and implementation techniques by:
        \begin{itemize}
            \item Identifying environmental conditions and types of degradation that can be utilized.
            \item Suggesting structural component designs that exploit degradation for various transformations and functions.
            \item Proposing sequential control methods that incorporate triggered degradation.
            \item Examining candidate materials for triggerable degradation and their evolving mechanical properties.
        \end{itemize}
    \item Showcasing application examples across diverse environmental conditions to demonstrate the utility of the \textcolor{black}{DtF} design strategy.
\end{itemize}

\section{Related Work}

\subsection{Sustainable HCI}
Sustainability has gained increasing attention in the field of HCI over the past few decades \cite{blevis_sustainable_2007, silberman_next_2014, knowles_design_2016, bremer_have_2022, Patel_2023, Sustainflatable}. This heightened focus arises from the recognition of technology's significant impact on various aspects of life, including environmental sustainability \cite{silberman_next_2014, chisig}, social equity \cite{mankoff_hci_2012}, and economic inclusivity \cite{knowles_exploring_2013}, among others. Several research studies have investigated easy-to-unmake techniques, including methods such as 3D printing \cite{song_unmaking_2021}, smart textiles \cite{wu_unfabricate_2020}, and material-driven practices \cite{murer_-crafting_2017}. Sustainability efforts also focus on upcycling via creative reuse \cite{hanton_fabricatink_2022, williams_upcycled_2021} and recycling by repurposing materials \cite{wall_scrappy_2021, bell_reclaym_2022, Dew_Designing_2019}, primarily addressing resource-related challenges. 

In addition, several strategies have focused on utilizing bio-based materials and their biodegradable properties to address waste issues. These innovations have led to various applications and devices, including prototyping sensors and electronics \cite{kikkawa_photo-triggered_2017, song_towards_2022, Song_2023, Guridi_2023, Jin_2016}, and advancements in soft actuators \cite{hartmann_becoming_2021, Pataranutaporn_Toward_2018, waxpaperactuator}. While these approaches tackle end-of-life issues and waste management, many devices still rely on non-biodegradable components such as batteries and semiconductors. Another exploration avenue is the development of fully self-contained devices made entirely from biodegradable materials. These initiatives explore emerging organic materials \cite{kan_organic_2017} such as agar-based bioplastic \cite{bell_designing_2022}, alginate-based bioplastic \cite{Mihaleva_Haikeus_2023}, gelatin-based biofoam \cite{Vasquez_Exploring_2022}, flour dough-derived clay \cite{buechley_3d_2023}, and compost \cite{bell_reclaym_2022}. They also involve the use of SCOBY biofilms \cite{bell_scoby_2023} and interactive devices based on mycelium \cite{genc_interactive_2022, Vasquez_From_2019, Weiler_Mycelium_2019}.

It's commonly considered that the functional phase, where the device is operating, and the end-of-life phase, where the device is degraded, are mutually exclusive. To ensure performance during the functional phase, degradation should commence only upon disposal. Hartmann et al. \cite{hartmann_becoming_2021} have suggested that this dilemma can be resolved through built-in degradation triggers like enzymes, heat, or light. These triggers can naturally occur in different environments or be artificially induced \cite{kikkawa_photo-triggered_2017}. Thinking from a different perspective, Functional Destruction \cite{cheng_functional_2023} proposes that in some cases, the material's destruction can be leveraged to achieve specific functional goals of transient electronics. \textcolor{black}{Vasquez et al. investigated how intentional dissolving of biofoam yarns can be leveraged for designing fashion wearables \cite{Vasquez_2023}}. 

\textcolor{black}{Inspired by these precedents, our exploration of the \textcolor{black}{DtF} concept involves reimagining the end-of-life phase of materials as a stage where the morphing \textcolor{black}{device} remains functionally active}. In this approach, we leverage temporal and spatial fluctuations in environmental conditions to initiate sequential degradation and activate various functionalities. Our entire device operates autonomously, devoid of any external power sources, and is constructed entirely from degradable natural materials.

\begin{figure*}[t]
  \includegraphics[width=1\linewidth]{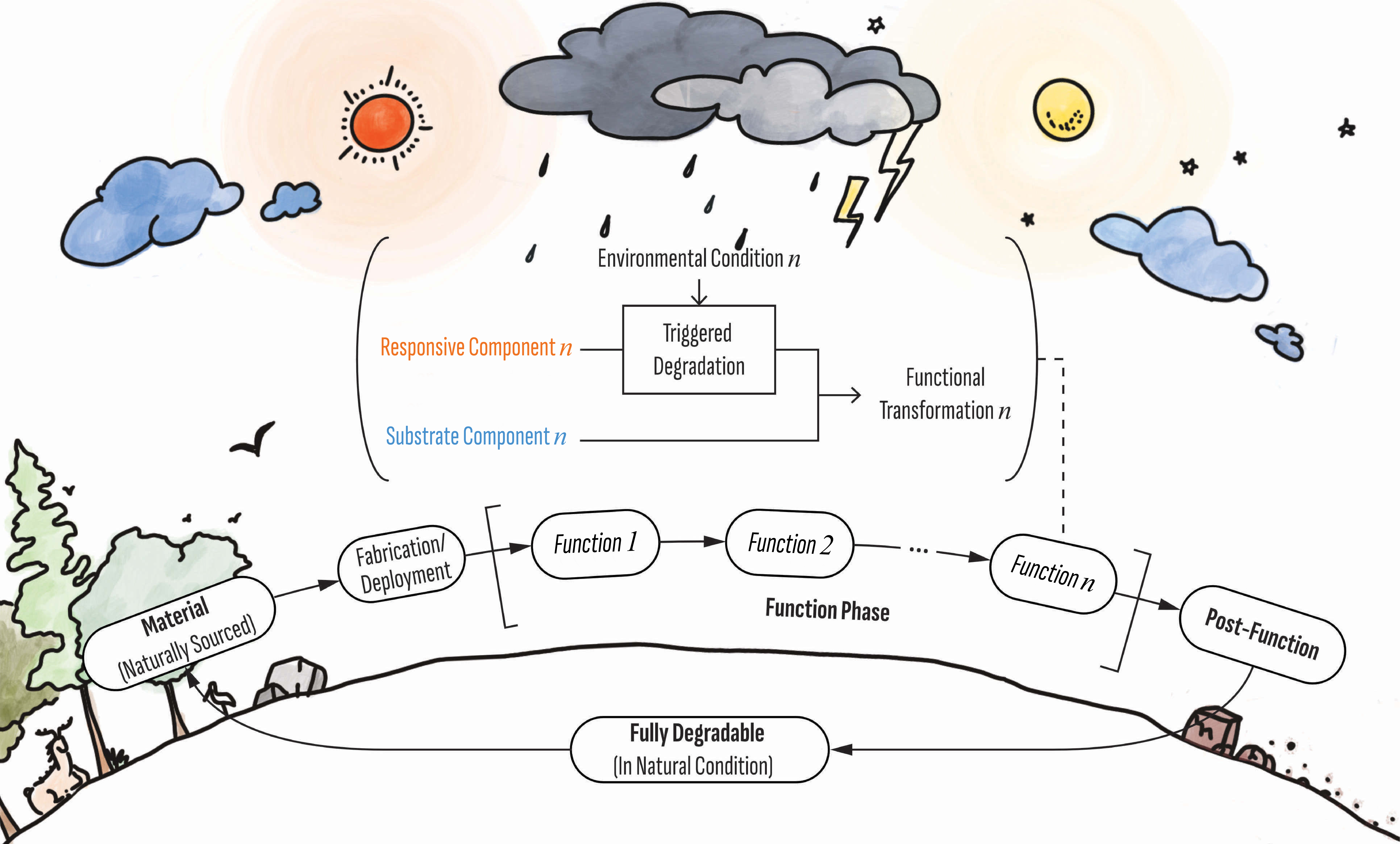}
  \caption{The “Degrade to \textcolor{black}{Function}” strategy overview.}
  \Description{}
  \label{fig:strategy}
\end{figure*}

\subsection{Morphing Devices Enabled by Responsive Material}
Cutting-edge advancements in morphing devices development have been driven by innovative materials that can react to a wide range of stimuli. Among these, heat-responsive materials have gained prominence in the creation of programmable morphing objects. Instances include the utilization of shape-memory alloys \cite{forman_modifiber_2019}, epoxy resin castings \cite{zhong_epomemory_2023}, 3D-printed polymers \cite{wang_printed_2018, wang_-line_2019}, and laminated bilayer films \cite{heibeck_unimorph_2015} to craft \textcolor{black}{devices} that react to changes in temperature. Additionally, colorimetric reagents \cite{kan_organic_2017, vega_dermal_2017} and synthetic hydrogels \cite{rizwan_ph_2017, hendi_healthcare_2020} have been developed to build \textcolor{black}{devices} that change color or serve as components in micro-scale robots for drug delivery, responding to shifts in pH levels. Researchers have also explored the potential of various materials, including films coated with bacteria \cite{yao_biologic_2015}, chemically-washed and molded wooden veneer \cite{luo_autonomous_2023}, and humidity-sensitive cholesteric liquid crystal networks (CLCNs) with magnetic composites \cite{feng_cholesteric_2023}. \textcolor{black}{Lastly, morphing devices that leverage the storage and release of elastic energy in materials have also been explored \cite{John, science, hawkes2022engineered}.}

These materials hold promise for creating interactive devices that are triggered by fluctuations in moisture levels. Similarly, materials that react to magnetic fields \cite{wakita_blob_2012}, electric fields \cite{lu_lime_2016, metalife}, and UV light \cite{jin_photo-chromeleon_2019} have witnessed extensive utilization as well.

However, many of the responsive materials mentioned above can be both costly and challenging to produce, often requiring specialized laboratory procedures, conditions, and equipment. This limits their widespread adoption and applicability. Additionally, these materials may necessitate specialized recycling processes, contributing to a high "embodied energy" and associated carbon dioxide emissions over their lifecycle \cite{lazaro_vasquez_introducing_2020}. In contrast, this paper focuses on using (commercially-available) \textcolor{black}{natural materials or composites made from organic, mineral, or other naturally occurring resources through simple processing methods like dissolving, melting, casting, and molding}. These materials are used to create the \textcolor{black}{DtF} morphing \textcolor{black}{devices} that can respond to various stimuli, making our approach both more accessible and sustainable.

\subsection{Beneficial Degradation}
\textcolor{black}{Although it might seem counterintuitive, the breakdown of materials within a functioning device can sometimes be beneficial.} For example, fuses are designed to blow when there is an overload, acting as a safety mechanism in electrical circuits \cite{EDISON_1890}. In marine and architectural applications, cathodic protection uses sacrificial anodes to mitigate the corrosion of metal surfaces \cite{pedeferri_cathodic_1996}. Similarly, helmets are \textcolor{black}{designed} to fracture under extreme stress, thereby absorbing and dissipating energy to protect the wearer \cite{fernandes_helmet_2019, bhudolia_enhanced_2021}. Laundry detergents and dishwasher capsules are designed to dissolve easily, allowing for efficient transportation of the liquid contents and reducing waste after use \cite{byrne_biodegradability_2021}.

\textcolor{black}{Nature also provides compelling examples of "purposeful" degradation. Certain plants, like the squirting cucumber \cite{galstyan_snap_2018}, oxalis \cite{li_seed_2020}, orange jewelweed \cite{hayashi_mechanics_2009}, and pletekan \cite{soetedjo_mechanical_2013}, have seed capsules that burst open to disperse seeds when structural integrity fails. The seed coat of species like black wattle cracks open due to the heat from fires, triggering germination to aid in post-fire ecological recovery \cite{tangney_success_2022}. Natural erosives have also crafted unique landscapes, such as Yardangs \cite{blackwelder_yardangs_1934} and Stalactites \cite{noauthor_stalactites_nodate}. These instances of degradation fulfill specific, interesting functions, inspiring us to design morphing devices that transform stimulus-triggered degradation into a beneficial function.}

\section{DtF Strategy Overview}

Degradation often negatively impacts the stability of materials and the function of an operating device. However, the \textcolor{black}{DtF} concept proposes using controlled, gradual degradation of materials and structures to initiate a beneficial transformation, or "\textcolor{black}{Function}". This adaptive transformation allows the device not only to adapt to changing environmental conditions, but also to have a positive impact on the environment itself. By strategically \textcolor{black}{leveraging} degradation, \textcolor{black}{DtF} introduces a novel way to imbue morphing devices with inherent adaptability and sustainability 
(Fig. \ref{fig:strategy}). 

\textcolor{black}{Materials may degrade at different rates, and their degradation behavior can be distinctively engineered in response to various environmental factors. } Such variability offers a unique opportunity to design dynamic morphing devices that experience \textcolor{black}{degradation in a controlled manner under specific conditions, a process we refer to as programmed degradation}. The result is a sequence of transformations, enabling the execution of diverse functions either in distinct phases or in response to specific environmental triggers. This approach also informs the design of self-sufficient devices that operate autonomously, eliminating the need for external power or manual interventions to trigger their functions. As a result, leveraging the inherent properties of materials for programmed degradation can yield designs that are not only adaptable but also self-reliant and environmentally responsive.

In order to systematically address the different elements in designing such a morphing \textcolor{black}{device}, we've classified its structural components into two categories based on their degradation rates: \textcolor{black}{\emph{substrate and responsive components}}. Substrate components, characterized by their slower degradation process across diverse environments, provide a fundamental support framework. This category encompasses components that function in elastic energy storage, stationary support, protection, and motion transmission, etc.

The responsive component is initially stable but specifically designed to degrade more rapidly under particular environmental conditions. Once the component failures occur due to this accelerated degradation, the device triggers a pre-designed transformation to fulfill its intended function. By carefully selecting materials and structural designs, it's possible to create devices that respond to a wide array of environmental stimuli, undergoing varied transformations and functionalities. Furthermore, the temporal and spatial fluctuations of environmental conditions can be utilized to orchestrate these transformations, tailored to the specific needs of the application.

In this paper, our proposition involves employing natural materials sourced from organic, mineral, or other naturally occurring resources. This category includes biosynthetic materials like wood and fibers, mineral-based substances like calcium carbonate (CaCO$_{3}$), and elemental metals such as magnesium (Mg). Such kinds of material are generally considered benign to their surrounding ecosystems during their lifecycle and do not introduce harmful substances into the environment upon degradation \textcolor{black}{\cite{Amurakimi_Biodegradable_2022, Dong_Environmentallay_2020, kuo_bioinspired_2018}}. 


\section{Design Considerations}
\label{section:design_consideration}
In this section, we detail the design factors that should be considered when developing a \textcolor{black}{DtF} morphing device. A general \textbf{design flow} would be: 1) Assessing the primary factors of the environmental conditions and potential fluctuations in the usage setting (Section \ref{sec:environmental_conditions}). 2) Specifying potential triggered degradation that can be leveraged (Section \ref{sec:degradation}). 3) Designing the device to achieve the desired sequential transformation through degradation (Section \ref{sec:component_design}, \ref{sec:sequential}). 4) Selecting materials aligned with the above points and implementing the device (Section \ref{section:material_selection}). 5) Testing and iteration.

\subsection{Environmental Conditions}
\label{sec:environmental_conditions}
\begin{figure}[b]
  \includegraphics[width=\columnwidth]{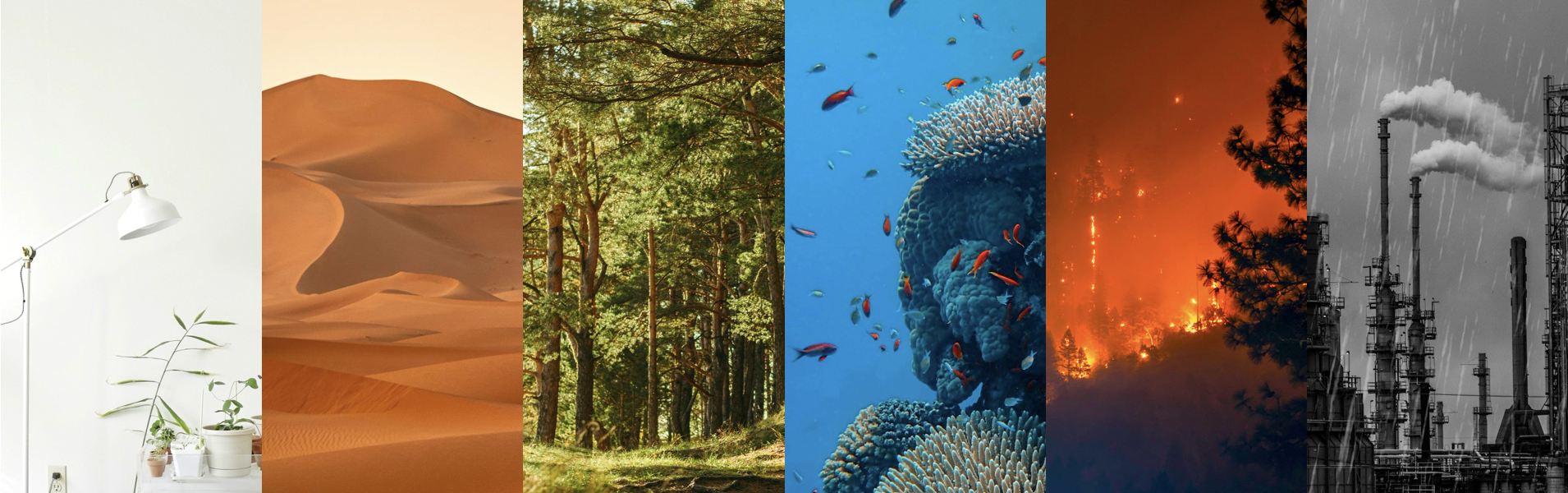}
  \caption{Several environmental condition examples.
  }
  \Description{}
  \label{fig:environment_condtion}
\end{figure}

Different spaces or terrains dictate distinct environmental conditions based on fundamental environmental factors. Here we summarized several \textcolor{black}{representative} conditions, emphasizing featured environmental factors of each (Fig. \ref{fig:environment_condtion}):

Room - A standard room often maintains 25$^\circ$C with a 30\%-50\% relative humidity (RH) \cite{noauthor_ideal_nodate}. Desert - Day temperatures vary between 30$^\circ$C to 50$^\circ$C, coupled with a 10\% to 30\% low RH, with occasional water sources or rainfall. \cite{desert_nasa, Waggoner}. Forest - Typical and temperate forests can be warm with RH levels exceeding 80\% \cite{rainforest_nasa, noauthor_tropical_2023}. Aquatic - Its defining characteristic is being submerged in water. Extreme - Includes scenarios like wildfires causing severe temperature spikes and land contamination leading to soil acidification with pH sometimes dropping below 4 \cite{thomas_soil_1996}. In addition, the microbes and macroorganisms are usually prevalent in many of these environments.


Furthermore, these environmental conditions are not static. For instance, desert temperatures can plummet at night, and temperate forests would only experience increased warmth and humidity as spring progresses. Spatial variations also occur: deserts may feature oases with small water bodies, and high-altitude mountain regions might possess microclimates varying from the surrounding areas.

When selecting materials for the responsive component of a \textcolor{black}{DtF} morphing \textcolor{black}{devices}, it's crucial to consider whether the degradation rate might be significantly influenced by targeted environmental conditions. Moreover, understanding the \emph{temporal and spatial condition changes} in these environments can inform decisions about the optimal timing and method for their degradation.


\subsection{Degradation}
\label{sec:degradation}
\begin{figure}[b]
  \includegraphics[width=\columnwidth]{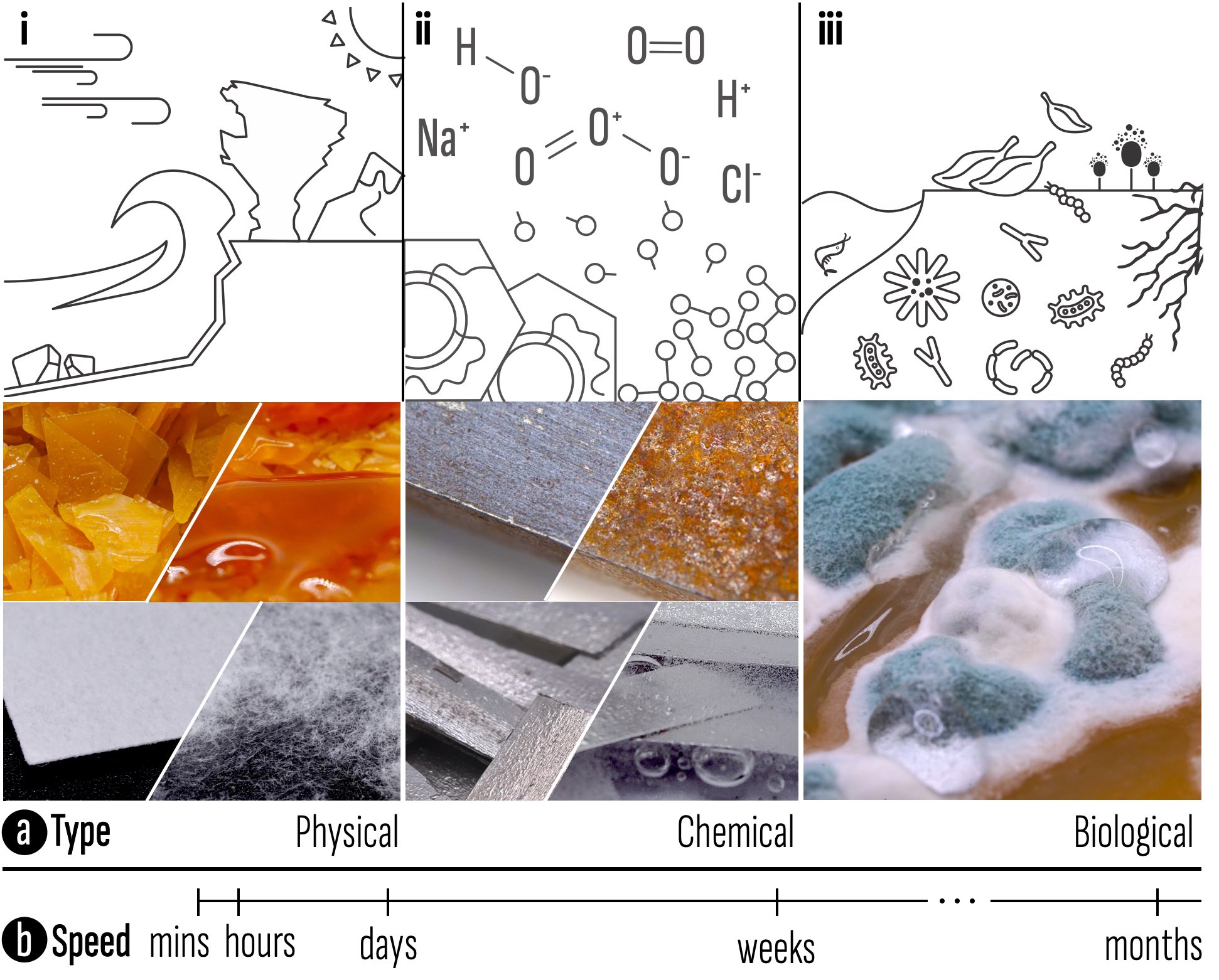}
  \caption{Degradation types and speed considered in \textcolor{black}{DtF}. a) Types: i. Physical (e.g., melting, dissolving); ii. Non-biological Chemical (e.g., rusting, corrosion); iii. Biological. (b) Speed.}
  \Description{}
  \label{fig:degradation}
\end{figure}

\textit{Degradation Types}. \textcolor{black}{Three primary types of degradation can be considered for leverage} (Fig. \ref{fig:degradation}.a). Physical degradation: Entails the disintegration of materials through processes such as melting, dissolving, erosion, and fragmentation without altering their chemical structure. Chemical degradation: Described here as the transformation of a material's molecular structure and properties through non-biological chemical reactions, such as metal rusting or acid corrosion. Biological degradation: The breakdown of materials into simpler compounds by living organisms including micro and macroorganisms, facilitated through enzymatic actions. 


\noindent\textit{Degradation Speed}. One of the pivotal aspects of constructing a \textcolor{black}{DtF} morphing \textcolor{black}{device} is leveraging the differential degradation speeds of materials under varying conditions. The guiding principles are  (Fig. \ref{fig:degradation}.b):
\begin{itemize}[leftmargin=10pt]
    \item Stability outside targeted conditions: Materials, when not subjected to the targeted conditions, should exhibit stability. This translates to a slow degradation speed, aligning with the desired shelf-life.
    \item Accelerated degradation in targeted conditions: The responsive component material should manifest a markedly accelerated degradation rate when exposed to the specific targeted conditions. In the current \textcolor{black}{DtF} morphing \textcolor{black}{devices}, we have designated a period of less than two weeks as a preferred timeframe.
    \item Substrate component longevity: The substrate component material should be durable enough to sustain until the conclusion of the function phase under targeted conditions.
\end{itemize}

\subsection{Structural Component Design}

\begin{figure}[b]
  \includegraphics[width=\columnwidth]{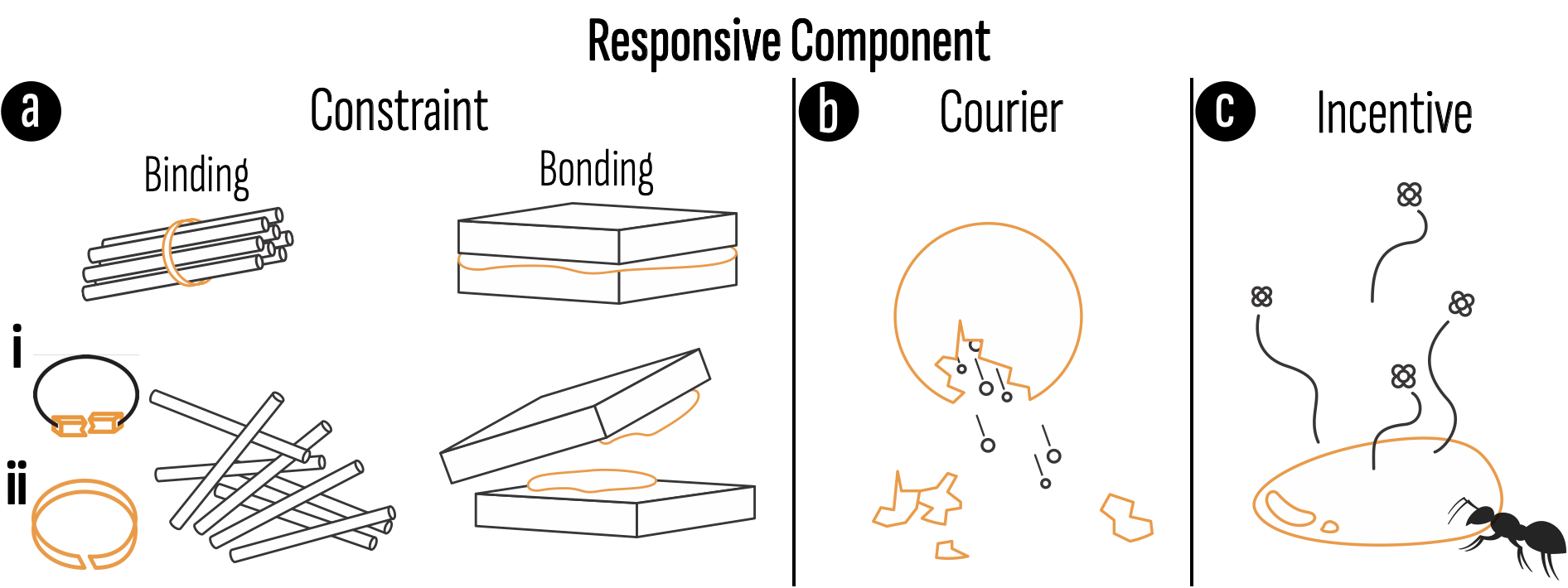}
  \caption{Three types of responsive components: a) constraint, b) courier, c) incentive}
  \Description{}
  \label{fig:responsive_structure}
\end{figure}

\setcounter{figure}{7}
\begin{figure*}[b]
   \centering
  \includegraphics[width=1\textwidth]{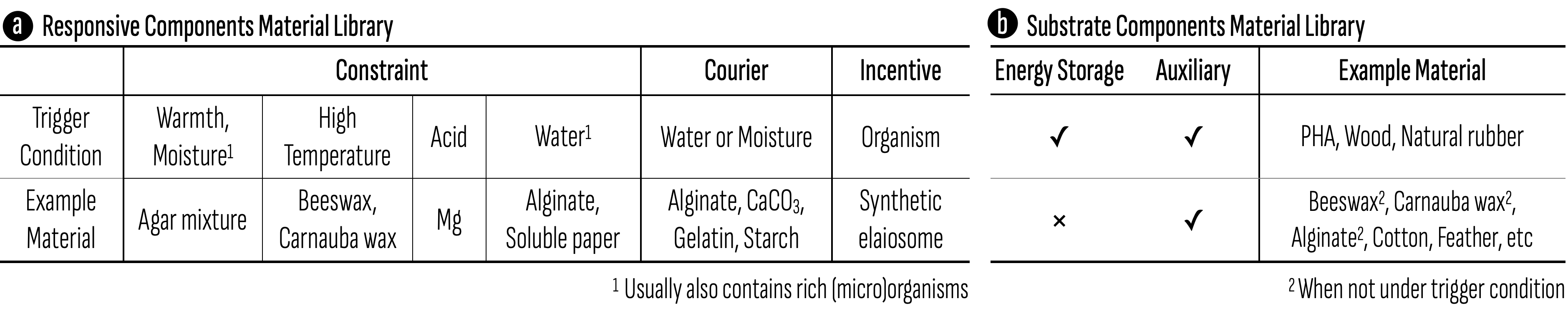}
  \caption{\textcolor{black}{The example materials for a) responsive and b) substrate components.}}
  \Description{}
  \label{fig:material_selection}
\end{figure*}

\label{sec:component_design}
\subsubsection{Responsive components} play a critical role in the \textcolor{black}{DtF} morphing \textcolor{black}{device}. They need to experience accelerated deterioration, primarily when they encounter specific environmental triggers. \textcolor{black}{We have divided such components into three categories (Fig. \ref{fig:responsive_structure}) based on their primary functions}:




\begin{itemize}[leftmargin=10pt]
    \item \textit{Constraint} restricts the motion or deformation of the substrate component. When the constraint degrades and loses its function, the substrate component can then be activated to achieve the desired locomotion or deformation. Constraints can be categorized into two primary types: binding and bonding, as shown in Fig. \ref{fig:responsive_structure}.a. Binding itself takes two forms. In the case of castable materials that are non-compliant when dry (e.g., agar, alginate), they serve as connectors for severed elements like cotton threads (Fig. \ref{fig:responsive_structure}.a.i). Other materials like paper or metal sheets may be used as ribbons (Fig. \ref{fig:responsive_structure}.a.ii). This allows for the creation of versatile and secure constraints. On the other hand, bonding generally involves using materials as adhesives to join components together. 
    \item \textit{Courier} could be a capsule crafted from responsive materials or a combination of the responsive material and the ingredients to be delivered. Once the courier degrades, it facilitates the release and delivery of its contents (Fig. \ref{fig:responsive_structure}.b). In this paper, we mainly focus on water-triggered delivering.
    \item \textit{Incentive}, serving as a "bait", can be specifically designed to attract particular organisms. The appeal of the incentive encourages these organisms to consume it, leading to the degradation of the component (Fig. \ref{fig:responsive_structure}.c).
\end{itemize}

\setcounter{figure}{5}
\begin{figure}[t]
  \includegraphics[width=\columnwidth]{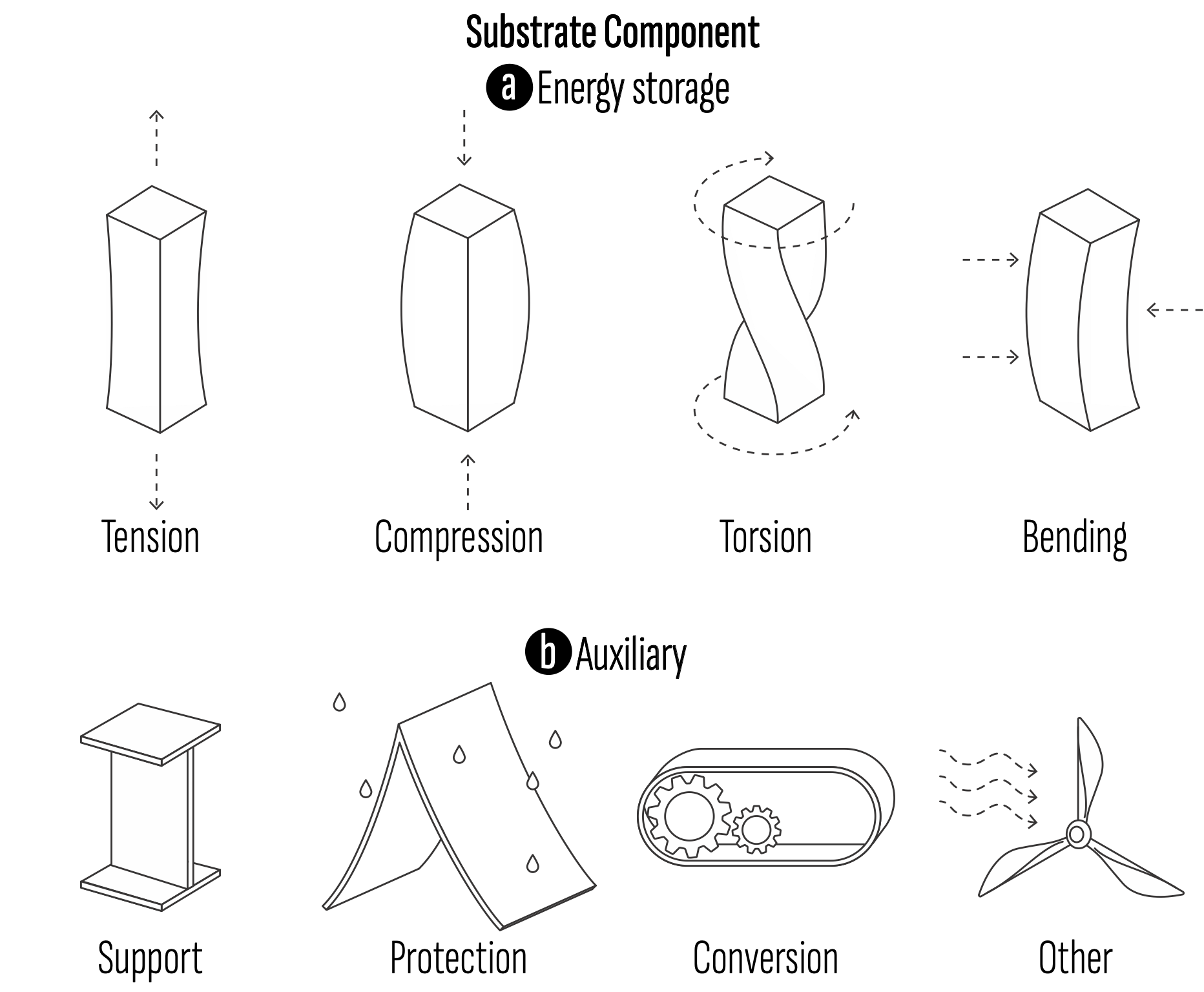}
  \caption{The substrate components: a) Energy storage, including tension, compression, torsion and bending; b) Auxiliary, including support, protection, energy conversion and other power components.}
  \Description{}
  \label{fig:substrate_structure}
\end{figure}

\subsubsection{Substrate components} They need to maintain relatively robust material stability and mechanical properties throughout the devices' functional phase. And they can be designed to be passively activated to perform specific functional transformations when the responsive components undergo accelerated degradation. \textcolor{black}{We divide substrate components into two categories (Fig. \ref{fig:substrate_structure})}:
\begin{itemize}[leftmargin=10pt]
    \item Energy Storage: These components are engineered to store elastic potential energy, such as through compression, tension, bending, or torsion (Fig. \ref{fig:substrate_structure}.a). They primarily work in conjunction with responsive constraint components. The stored energy is unleashed when the constraint degrades in response to environmental triggers, enabling the energy storage component to revert to its original form and execute its intended function.
    \item Auxiliary: \textcolor{black}{These elements provide various types of foundational support, such as stands, beams, and joints (Fig. \ref{fig:substrate_structure}.b).} The protection is designed to shield the device from undesirable environmental factors, such as a rain shelter. The conversion transfers or converts mechanical energy, such as belts and pulleys. Other power components are mechanisms augment force or power for specialized functions like buoyancy devices and wings.
    
\end{itemize}

\subsection{Sequential Transformation}
\label{sec:sequential}
In this section, we explain how triggered degradation can be harnessed to achieve sequential transformations. Transformations predominantly occur when the responsive constraint component fails, thereby releasing the elastic energy within the energy storage component. We propose three basic types of sequential transformations that can be achieved by leveraging temporal and spatial changes in environmental conditions. These basic types of sequential transformations can also be combined for more complex behaviors.

\begin{figure}[t]
  \includegraphics[width=\columnwidth]{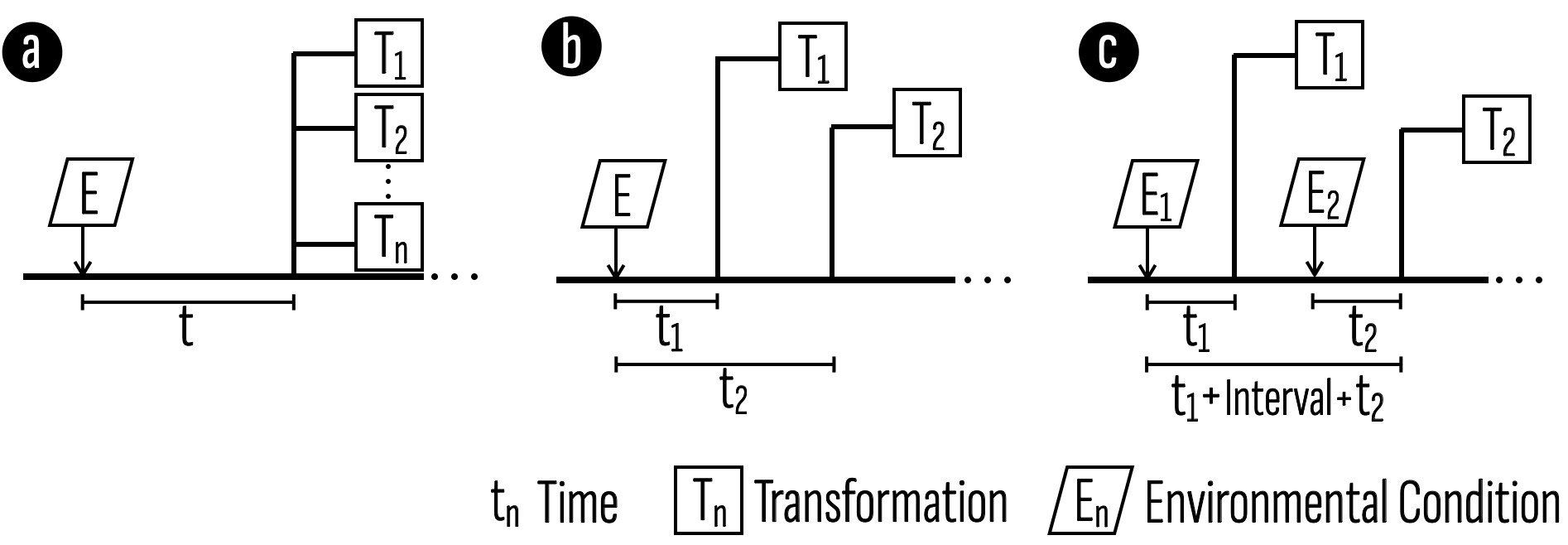}
  \caption{Three basic approaches to achieve sequential transformations in a \textcolor{black}{DtF} morphing \textcolor{black}{device}.}
  \Description{}
  \label{fig:sequence}
\end{figure}

\begin{itemize}[leftmargin=10pt]
    \item An environmental condition triggers one or multiple constraints to degrade and fail nearly simultaneously, initiating several transformations at the same time (Fig. \ref{fig:sequence}.a).
    \item An environmental condition triggers multiple constraints to degrade and fail at distinct times, enabling asynchronous transformations to occur (Fig. \ref{fig:sequence}.b).
    \item An environmental condition triggers a constraint to degrade and release a transformation; subsequent new environmental condition lead to the failure and release of another constraint and transformation (Fig. \ref{fig:sequence}.c). 
\end{itemize}

\subsection{Material Selection}
\label{section:material_selection}

\setcounter{figure}{8}
\begin{figure*}[b]
   \centering
   \includegraphics[width=1\textwidth]{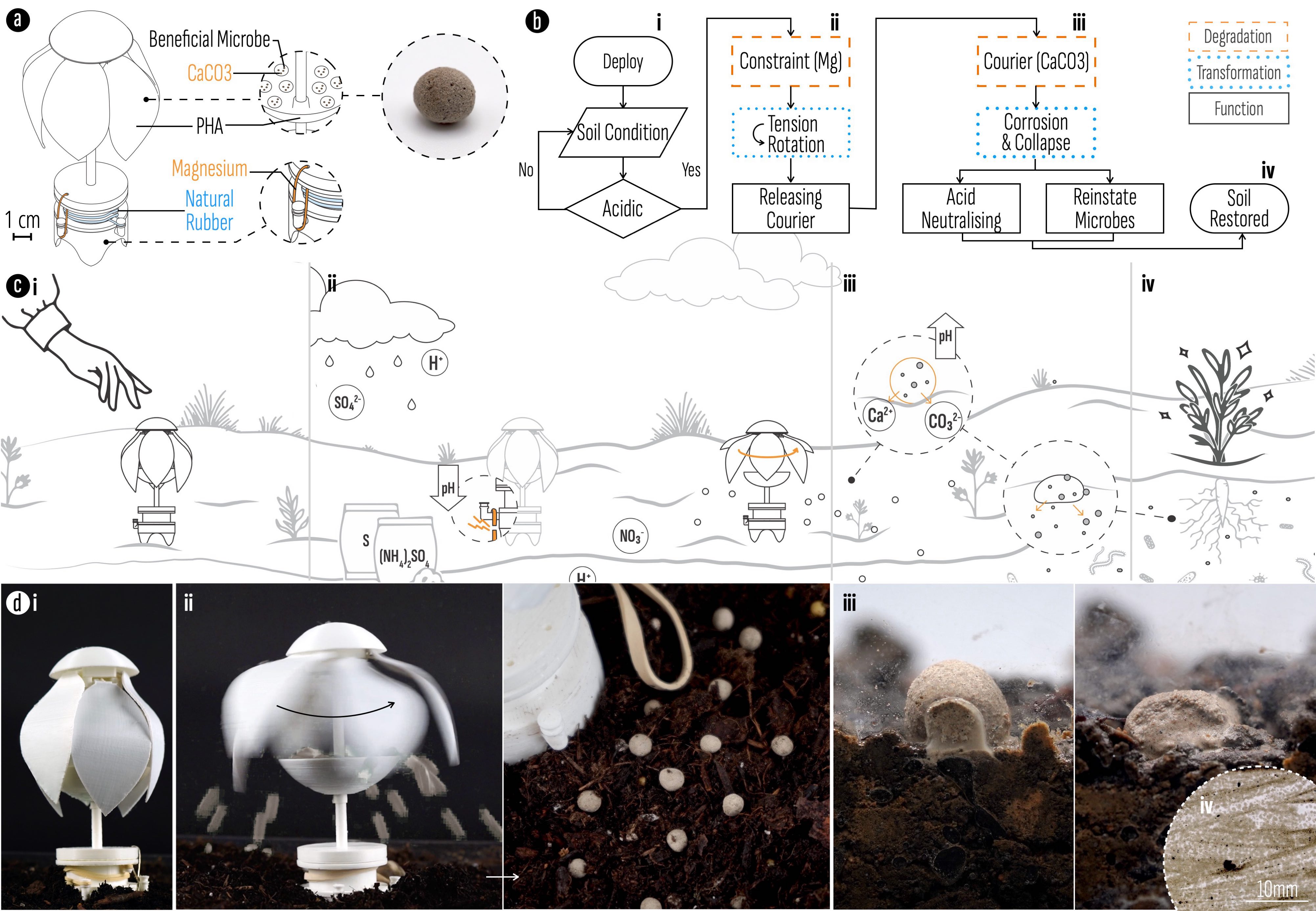}
   \caption{The acidic soil monitoring and rehabilitation. a) Structure and material. b, c) The workflow: i. These devices can be deployed in areas at high risk for acidic pollution; ii. CaCO$_3$ couriers are released when acidic soil is detected; iii. The couriers neutralize the soil and release microbes; iv. Microbes restored.}
   \Description{}
   \label{fig:app_soil}
\end{figure*}

Here we summarize the primary principles for material selection in construction as follows: 1) it is sourced from natural origins, 2) it exhibits a considerable shelf-life when not deployed and properly stored, and 3) it is capable of natural degradation in the post-functional phase. 

\textcolor{black}{In addition, the materials used for the responsive component must maintain} sufficient tensile strength when utilized as a constraint, and exhibit stability under non-targeted environmental conditions while being responsive to environmental triggers. Meanwhile, materials chosen for the substrate component should ensure continued stability throughout the functional period. If for energy storage purpose, they should possess a high elastic modulus to facilitate significant potential for elastic energy storage without incurring permanent deformation. Moreover, excellent resilience is also required, which allows the material to revert to its original form with minimal energy loss, alongside sufficient yield strength to withstand the required stress before undergoing permanent deformation.

In Fig. \ref{fig:material_selection}, we present a reference material library, detailing the components each material is suited for and the environmental triggers to which the material responds. These materials have been selected based on the principles discussed above and through experimental validation. The detailed selection process will be discussed in Section \ref{sec:implementation_techniques}.

\begin{figure*}[b]
  \includegraphics[width=1\linewidth]{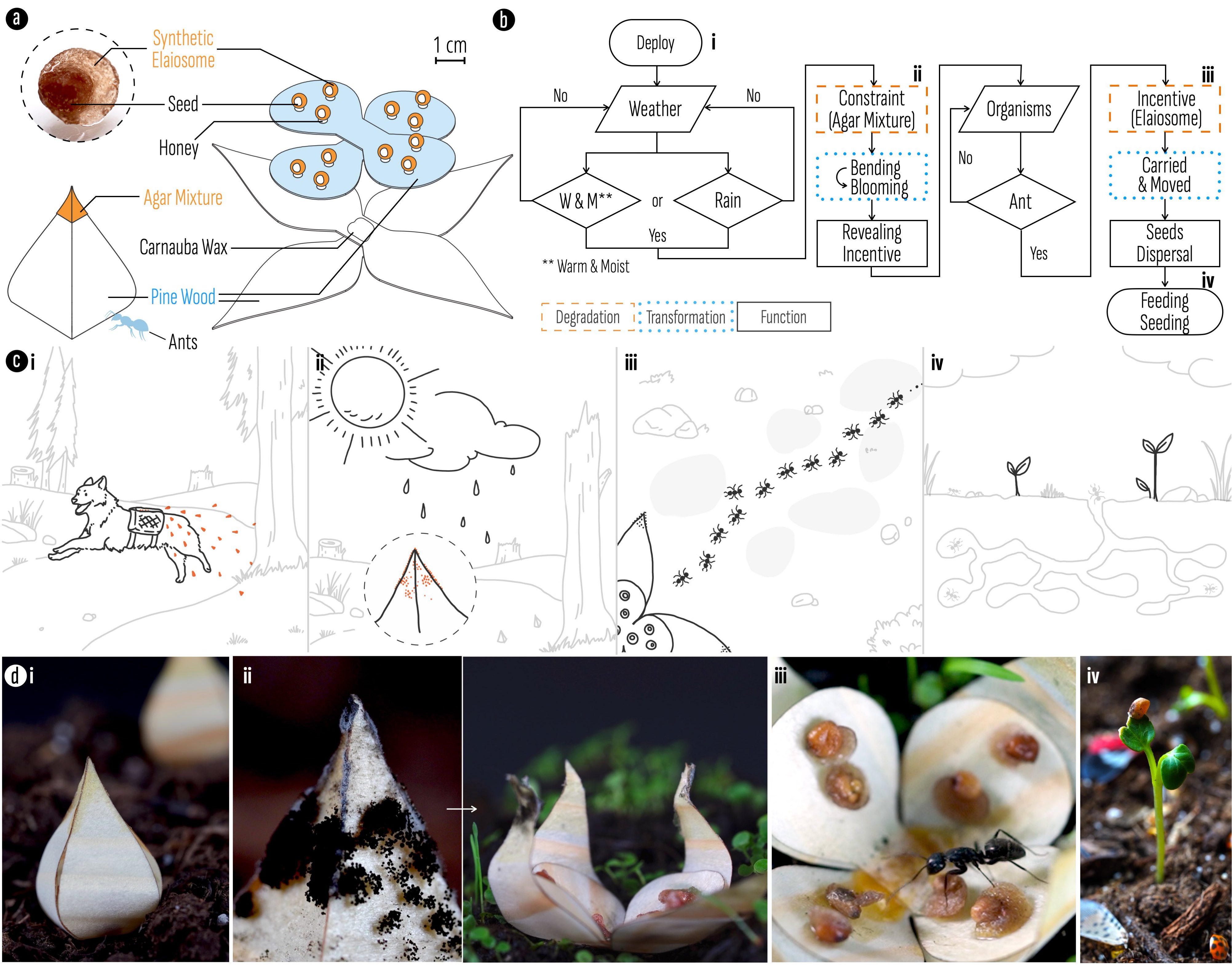}
  \caption{The forest feeder and seeder. a) Structure and material. b-d) The workflow: i. The pods may be distributed manually or with help of animals; ii. Molds begin to grow on the agar mixture constraint as the climate becomes warmer and more humid. Within a week or two, the constraint fails which allows the pod to gradually open; iii. Ants are attracted by the elaiosome coating and carry the seeds back to their nest; iv. After the elaiosome is consumed, the seeds may potentially be discarded in the nutrient-rich waste areas, providing an ideal environment for them to take root and germinate.}
  \Description{}
  \label{fig:app_ant}
\end{figure*}

\section{DtF Morphing Device Examples}
\label{sec:application}

\textcolor{black}{Engineered for autonomous and sustainable operation with naturally degradable materials, the DtF morphing device is especially suitable for outdoor environments, minimizing the need for ongoing intervention or external power sources. This eco-friendly feature inspired us to fully leverage its advantages and construct five applications targeted at restoration and rehabilitation, addressing real-world environmental issues. The multiple sequential transformations offer broad applicability and the ability to execute various functions to help achieve these goals.}

Each application is developed for a different type of environmental conditions, collectively encompassing every design considerations to exemplify DtF device design and inspire innovation among researchers and designers. Due to space limitations, three examples are included in this section. The other two, focusing on \textbf{desertification governance} and \textbf{wildfire monitoring}, and utilizing soluble paper and wax for responsive components, can be found in the \href{https://doi.org/10.1145/3654777.3676464}{\underline{Supplementary Material}}.

\textcolor{black}{Additionally, these are speculative applications validated in lab settings. For real-world deployments, it is crucial to acknowledge the necessity of further field tests and environmental deliberations. Specifically, such devices need to be distributed strategically, properly retrieved when necessary, and be cautious about introducing foreign materials. Long-term monitoring and assessment are also essential.} More discussions can be found in Section \ref{Discussion}.

\subsection{Acidic Soil Monitoring and Rehabilitation}


Soil health refers to the soil's ability to sustain life, regulate water, filter pollution, cycle nutrients, and provide physical support for plants, animals, and humans \cite{soil_health_2023b}. In natural settings, soil acidification advances gradually, but agricultural or gardening activities expedite this process through persistent utilization of sulfuric and ammonium-based fertilizers in conjunction with acidic precipitation \cite{goulding_soil_2016}. This acidity, which can plummet to a pH of 4 or even lower, significantly impacts the availability of nutrients to plants and disrupts the symbiotic relationships between microorganisms such as bacteria, fungi, and other microbes \cite{thomas_soil_1996, singh_acid_2008}. 

To address this challenge, we have \textcolor{black}{developed} a \textcolor{black}{DtF} device designed to continuously monitor soil pH and initiate corrective actions for soil rehabilitation. As illustrated in Fig. \ref{fig:app_soil}, these devices can be strategically deployed in areas at high risk for acidic pollution. After deployment, a responsive constraint component made of Mg foil serves as a pH sensor (Fig. \ref{fig:app_soil}.i). Should the soil turn acidic, the Mg will degrade much more rapidly—the rate of degradation increasing with lower pH levels. 
To ensure a timely response, we've calibrated the maximum tensile force the Mg constraint can withstand to be just about 20\% higher than the actual load ($\sim 8$ N). We conducted five tests, adding 0.05 mL of pH4 sulfuric acid (H$_2$SO$_4$) solution to Mg at a rate of once per second. The results showed constraints failed between 3 to 8 hours. Each failure occurred instantaneously.

Upon degradation, the Mg constraint fails, prompting a stretched rubber ring to contract and release its stored elastic energy. A belt and pulley mechanism then converts this linear motion into rotational motion, generating enough centrifugal force to open staggered, petal-like structures and disperse the CaCO$_3$ couriers (Fig. \ref{fig:app_soil}.ii). These couriers are designed to gradually degrade through a combination of water solubility and acid-base reactions, effectively neutralizing soil acidity \cite{mclean_soil_1983} (Fig. \ref{fig:app_soil}.iii). Additionally, encapsulated beneficial microbes (Raw Microbes Root, NPK Industries) are released to replace those killed off by soil acidification (Fig. \ref{fig:app_soil}.iv).

The device can be reset for further use. Its main substrate component, made from PHA, can remain functional for months before biodegrading. Considering that Mg is an essential nutrient for plants, and that soil acidification can result in a Mg deficiency in the soil, the small amount of Mg used in the constraint—approx. 6 mg—could be safely left in the soil without causing harm. \textcolor{black}{Furthermore, the Mg constraints can potentially be geometrically tuned to realize constraint failures occurring in various pH levels at desired time points. Additionally, this approach showed promising results in neutralizing acidic soil and restoring microbes in our preliminary experiments. Detailed results can be found in Appendix.\ref{appendix:Application}}. 

\subsection{Forest Feeder and Seeder}

\begin{figure*}[t]
  \includegraphics[width=1\linewidth]{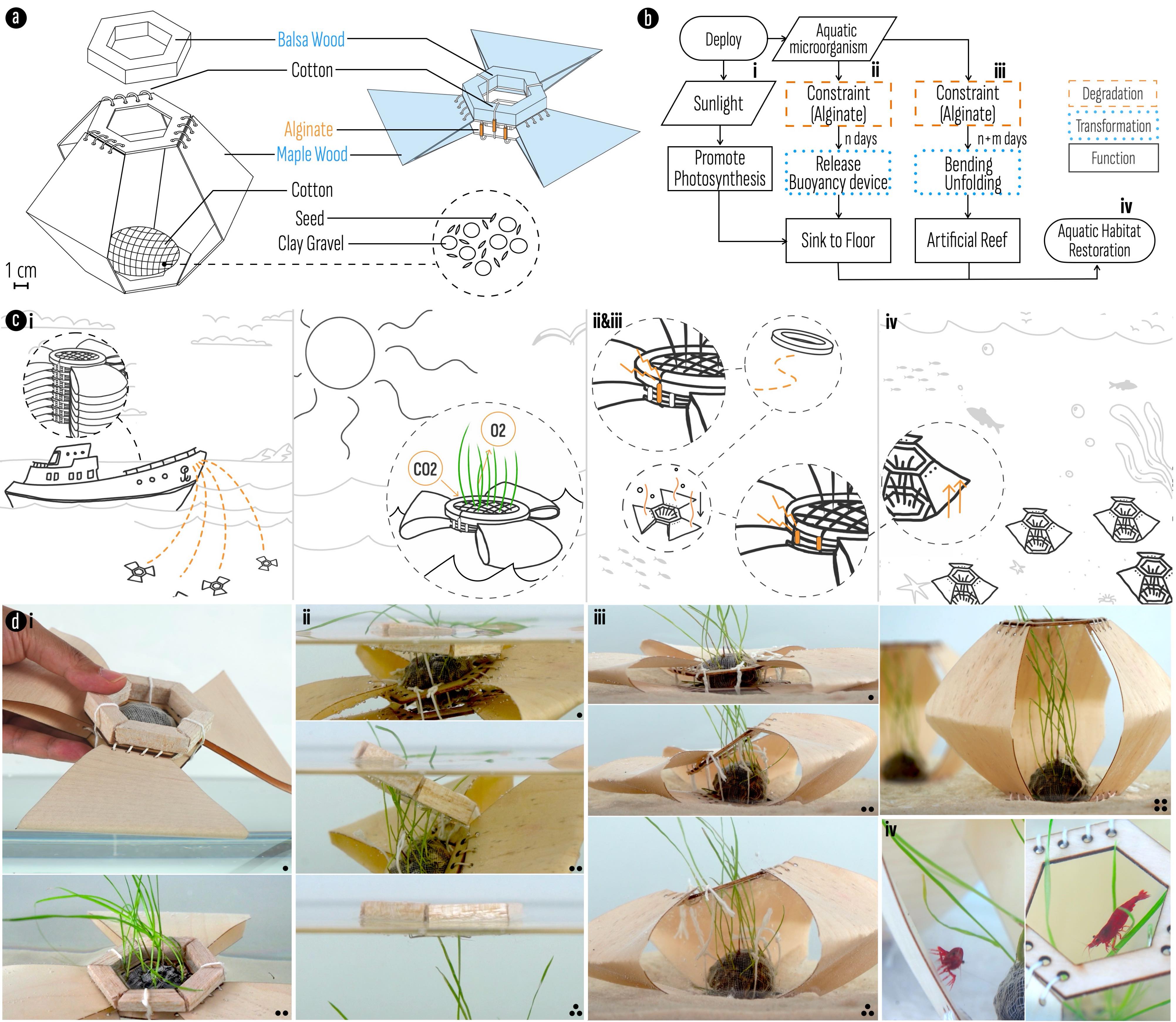}
  \caption{The Aquatic Habitat Restoration. a) Structure and material. b-d) The workflow: i. The artificial reef may be distributed to areas that need restoration. A buoyancy device keeps them afloat, allowing the seeds to germinate and flourish under abundant sunlight; ii. The first group of alginate constraints fail under microbes degradation. The buoyancy device detaches and the reef sinks; iii. As the second group of alginate layers fails, it unfolds to reveal the reef. This delay provides an opportunity for the grass to anchor itself to the ocean floor; iv. Shelters for small marine creatures were established.}
  \Description{}
  \label{fig:app_ocean}
\end{figure*}

\textcolor{black}{Plants employ various strategies for seed dispersal. One example is fleshy structures called Elaiosomes \cite{rcannon992_2023}, rich in fat and protein, attached to seeds to attract ants. Ants collect these seeds to nests, consume the Elaiosomes, and discard the seeds in nutrient-rich waste areas, aiding germination \cite{miller_effects_2020, sasidharan_seed_2019}. This symbiotic relationship benefits both ants, gaining nutrition, and seeds, dispersed to fertile grounds. This mechanism minimizes resource competition in the original location. Due to the critical role of seed dispersal in forest restoration and conservation, efforts have included using animals to help spread seeds in areas impacted by destruction. For example, dogs were employed to distribute native plant seeds in forests affected by burning \cite{Kaplan_2019}.}

Expanding on these revelations, we developed an automatic feeder and sower device. After being deployed in the forest, the pod lies in wait for the optimal moment to open (Fig. \ref{fig:app_ant}.i). As the climate becomes warmer and more humid, dormant mold spores activate and begin to grow on the agar mixture constraint that binds the pod  (Fig. \ref{fig:app_ant}.ii). Tests on five samples have shown that, within 6 - 11 days, the holding capacity of the agar mixture significantly decreases due to degradation, allowing the pod to open. This reveals seeds encased in an enticing synthetic elaiosome (s-eliosome, made from honey, peanut oil, agar, protein powder, and water) that the ants find irresistible (Fig. \ref{fig:app_ant}.iii). Drawn by the delicious morsel, ants carry the seeds back to their nests. There, the s-elaiosome is consumed, and the seeds may subsequently be discarded in the ants' nutrient-rich waste areas, providing an ideal environment for them to take root and germinate  (Fig. \ref{fig:app_ant}.iv). The oil content in the s-elaiosome may act as a moisture barrier, preventing premature seed germination until desired. We conducted preliminary tests in the field using seeds coated with s-elaiosome. We observed ants transporting the seeds and noted the presence of s-elaiosome-removed seeds around the nests. Continued observation can be conducted in the future to fully understand the dynamics of ant-seed interactions throughout the entire process.

The pod features a dual-layer structure crafted from finely cut dry pine wood veneer  (Fig. \ref{fig:app_ant}.a). The inner layer is designed to reinforce the opening of the outer layer once its constraints degrade. Both layers consist of two segments, arranged in a longitudinal petal-like configuration. The outer layer is cut to align with the wood grain, facilitating easier bending and shaping. In contrast, the inner layer is cut perpendicular to the wood grain to store more bending energy, increasing the pressure exerted on the outer layer. This amplifies the force applied to the agar mixture constraint by $\sim$ 5 times. These two layers are bonded with carnauba wax, while agar mixture secures the pod's tip, ensuring overall structural stability. The seeds are coated with s-elaiosome, and a small amount of honey improves their adhesion to the inner wall. All materials in this design are biodegradable and chosen for their environmentally-friendly decomposition characteristics.

\subsection{Aquatic Habitat Restoration}
\textcolor{black}{Seagrass and rivergrass meadows are vital in aquatic ecosystems, providing nourishment, habitats, and nursery grounds for various species \cite{short_global_2007, duffy_biodiversity_2015, moksnes_local_2018, Minguy_2021}. Despite their importance, these meadows have been steadily disappearing globally in recent decades \cite{orth_global_2006}. Understanding the life cycle and specific requirements of aquatic plants is crucial for successful restoration efforts.
Some underwater flowering plants, such as Posidonia oceanica, rely on photosynthesis for germination \cite{guerrero-meseguer_understanding_2018}. However, only about $\sim$45\% of solar energy penetrates to a depth of 1 meter underwater \cite{webb_65_nodate}. Moreover, tidal fluctuations and water currents pose challenges for seeds to take root in seabeds. To address these obstacles, researchers have explored using natural fiber bags for seed deployment \cite{unsworth_sowing_2019} and creating artificial reefs \cite{arredondo_blue_2022}.}

We've \textcolor{black}{developed an artificial reef} designed for delayed settling and autonomous deployment (Fig. \ref{fig:app_ocean}.a). Initially configured in a flat shape, these reefs allow for efficient packing and transportation. Once placed on the water's surface, a buoyancy device keeps them afloat, allowing the seeds within the seed pods to germinate and flourish under abundant sunlight (Fig. \ref{fig:app_ocean}.i). With five sample tested, the alginate constraints stand $\sim$ 3 N force. As alginate degrades based on the microbial content of the water, the first group typically fails within 6 to 10 days. At this point, the buoyancy device detaches, causing the reef to sink (Fig. \ref{fig:app_ocean}.ii). Utilizing an increased quantity of alginate constraint, double the amount compared to those that failed in the sinking phase, enables the reef to sustain itself for additional $\sim$ 5 to 11 days before unfolding (Fig. \ref{fig:app_ocean}.iii). This added delay may provide extra time for the grass to root and anchor to the ocean floor prior to reef expansion. Once unfolded, the expanded reef structure can potentially offer shelter to small marine creatures (Fig. \ref{fig:app_ocean}.iv). \textcolor{black}{We conducted experiments that demonstrate the promising effectiveness of this approach in promoting aquatic plant growth. Detailed results can be found in Appendix \ref{appendix:Application}.}

\begin{figure*}[t]
  \includegraphics[width=0.95\linewidth]{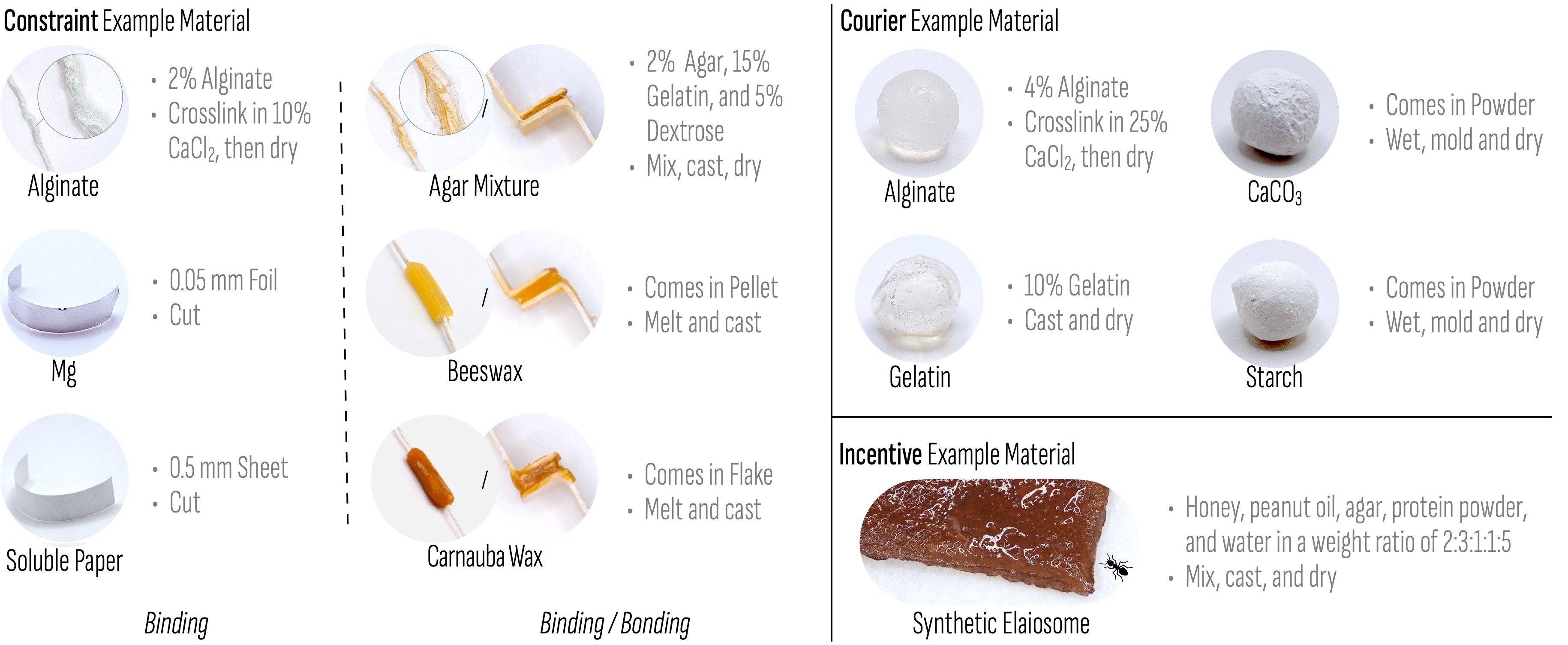}
  \caption{Exemplary materials suitable for each type of responsive component. The concentration unite is w/v (e.g, 2\% w/v Alginate). More detailed synthesis recipes, sources and procedures can be found at Appendix.}
  \Description{}
  \label{fig:responsive_material_1}
\end{figure*}

\section{Implementation Techniques}
\label{sec:implementation_techniques}
\subsection{Responsive Component and Material}\label{sec:responsive}

In Fig. \ref{fig:responsive_material_1}, we present exemplary materials suitable for each type of responsive component. For more detailed synthesis recipes and procedures, please refer to the Appendix \ref{appendix:A}, \ref{appendix:B}, \ref{appendix:C}. In the following sections, we delve into the specifics of their selection process.


\subsubsection{Constraint}
Traditional degradation studies often focus primarily on weight loss as the parameter of interest. However, for our application where the constraint component is key, it is critical to focus on changes in mechanical properties due to degradation, particularly tensile strength ($\sigma$). We define a term called Relative Tensile Strength ($\sigma_r$) to assess the material's capability during its degradation phase. The formula for calculating $\sigma_r$ is given by Equ. \eqref{eq:rel_tensile}, where "Maximum Load" represents the peak load the material can withstand, and $A_0$ denotes the original cross-sectional area of the material.
\begin{equation}
    \sigma_r = \frac{\text{Maximum Load}}{A_0}  \label{eq:rel_tensile}
\end{equation}

Based on previous discussions, we identify six environmental conditions to perform controlled experiments for material screening: 
1. 40\% RH at 25$^\circ$C (Room Condition),
2. 80\% RH at 30$^\circ$C (Forest-like Condition),
3. 20\% RH at 40$^\circ$C (Desert-like Condition),
4. 0\% RH at 70$^\circ$C (Wildfire-induced Condition),
5. Underwater with pH 4 at 25$^\circ$C (Acidification Pollution-induced Condition),
6. Underwater with microbes at 25$^\circ$C (Aquatic-like Condition).
For conditions 1-4, a combination of a controller, heater, and humidifier was used to regulate the RH and temperature inside a container. For condition 5, a solution was prepared using  sulfuric acid (H$_{2}$SO$_{4}$) and DI water. For condition 6, we used freshwater collected from a healthy aquarium system. Detailed experimental setup is outlined in Appendix \ref{appendix:A1}.

In total, \textcolor{black}{16} materials were tested. Depending on \textcolor{black}{the} materials’ initial forms, test samples were prepared either by cutting or casting. Casted samples were dried using a fan in ambient room conditions before undergoing any tests. Detailed information on sample dimensions and preparation procedures can be found in Appendix \ref{appendix:A2}.

Firstly, the initial tensile strength ($\sigma_i$) of each material was tested using five samples per material. Subsequently, 30 samples of each material were allocated to six environmental conditions. At specific intervals—every 1 minute (0-10 minutes), every 5 minutes (10-60 minutes), every hour (1-12 hours), and every day (1-14 days)—five samples for each material were taken out for tensile strength testing. If any of these five samples falls below predefined percentages of the $\sigma_i$ (80\%, 60\%, 40\%, 20\%, 10\%, 5\%), the $\sigma_r$ of all five samples was measured. Each percentage level was documented only once. For example, if a sample's $\sigma_r$ falls between 60\% and 80\% of $\sigma_i$, the next test will focus on whether any samples fall below 60\%. Samples that do not meet the criteria were returned to their respective environments. Additionally, long-term tests continue past the 14th day for the room condition group, with checks on the 30th, 60th, and 90th days.

\begin{figure*}
  \includegraphics[width=1\linewidth]{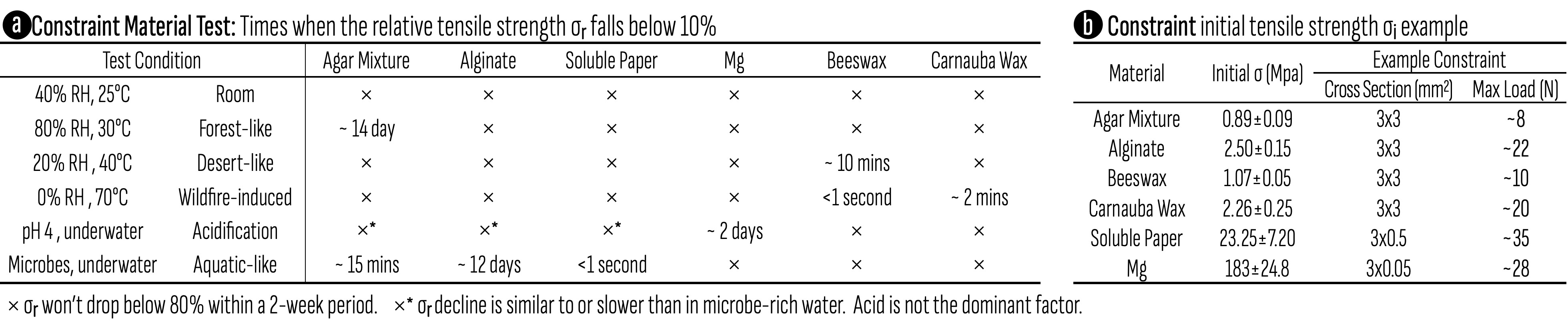}
  \caption{\textcolor{black}{a) Selected constraint material degradation test results. b) Example constraint tensile strength.}}
  \Description{}
  \label{fig:responsive_material_2}
\end{figure*}

Based on the guiding principle (Section \ref{section:material_selection}) and experimental result, we have selected six materials for the constraint component, with results presented in Fig. \ref{fig:responsive_material_2}.a, b. Detailed decline of the tensile properties over time, along with additional results for other non-selected candidate materials, are provided in in Appendix \ref{appendix:A3}. In summary:

\begin{itemize}[leftmargin=10pt]
    \item All selected materials exhibit an $\sigma_i$ around or exceeding 1 MPa.
    \item In room conditions, all materials demonstrate stability over a three-month duration, with their $\sigma_r$ remaining above 80\% of their $\sigma_i$.
    \item Each material is naturally sourced and is fully degradable post-functionality.
\end{itemize}

In addition, in triggering environmental conditions:

\begin{itemize}[leftmargin=10pt]
    \item Agar Mixture degrades significantly at 80\% RH and 30$^\circ$C. Early-stage $\sigma_r$ decline is primarily due to moisture absorption, while later-stage decline results from microbial growth and decomposition. Besides, it also exhibits degradation upon submersion in water.
    \item Alginate shows accelerated degradation underwater due to water absorption and microbial activity. However, In an 80\% RH, 30$^\circ$C environment, its $\sigma_r$ drops to around 80\% but stabilizes thereafter as non-aquatic microbes often cannot degrade alginate.
    \item Beeswax degrades faster at temperatures of 40$^\circ$C and 70$^\circ$C.
    \item Carnauba wax similarly degrades rapidly at 70$^\circ$C.
    \item Soluble Paper (0.5 mm thickness) disintegrates in seconds upon water contact.
    \item Mg (0.05 mm foil) remains relatively stable in freshwater but corrodes rapidly in pH4 solutions.
\end{itemize}

Among these, Mg stands out as it is a metal, yet it is considered a biodegradable metal and an essential nutrient. Like other inorganic materials such as Calcium (Ca), Iron (Fe), and CaCO$_3$, it forms part of the typical diet for living organisms and is found naturally in their composition \cite{vormann_magnesium_2003}. Given its high tensile strength, even a thin foil of Mg can serve effectively as a constraint, minimizing the amount needed.

\begin{figure}[b]
  \includegraphics[width=\columnwidth]{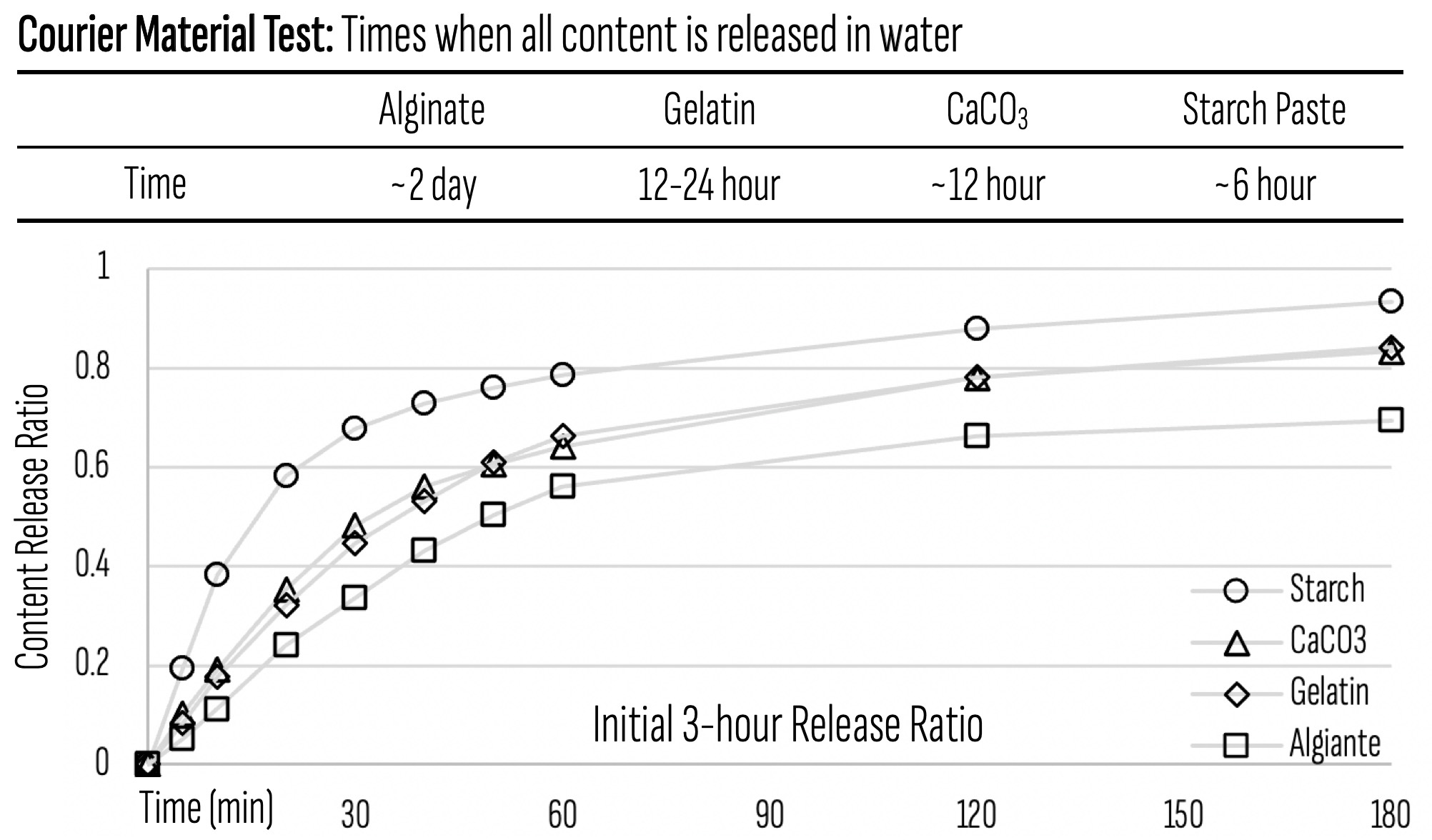}
  \caption{\textcolor{black}{Courier material test results.}}
  \Description{}
  \label{fig:responsive_material_3}
\end{figure}

\begin{figure}[b]
  \includegraphics[width=\columnwidth]{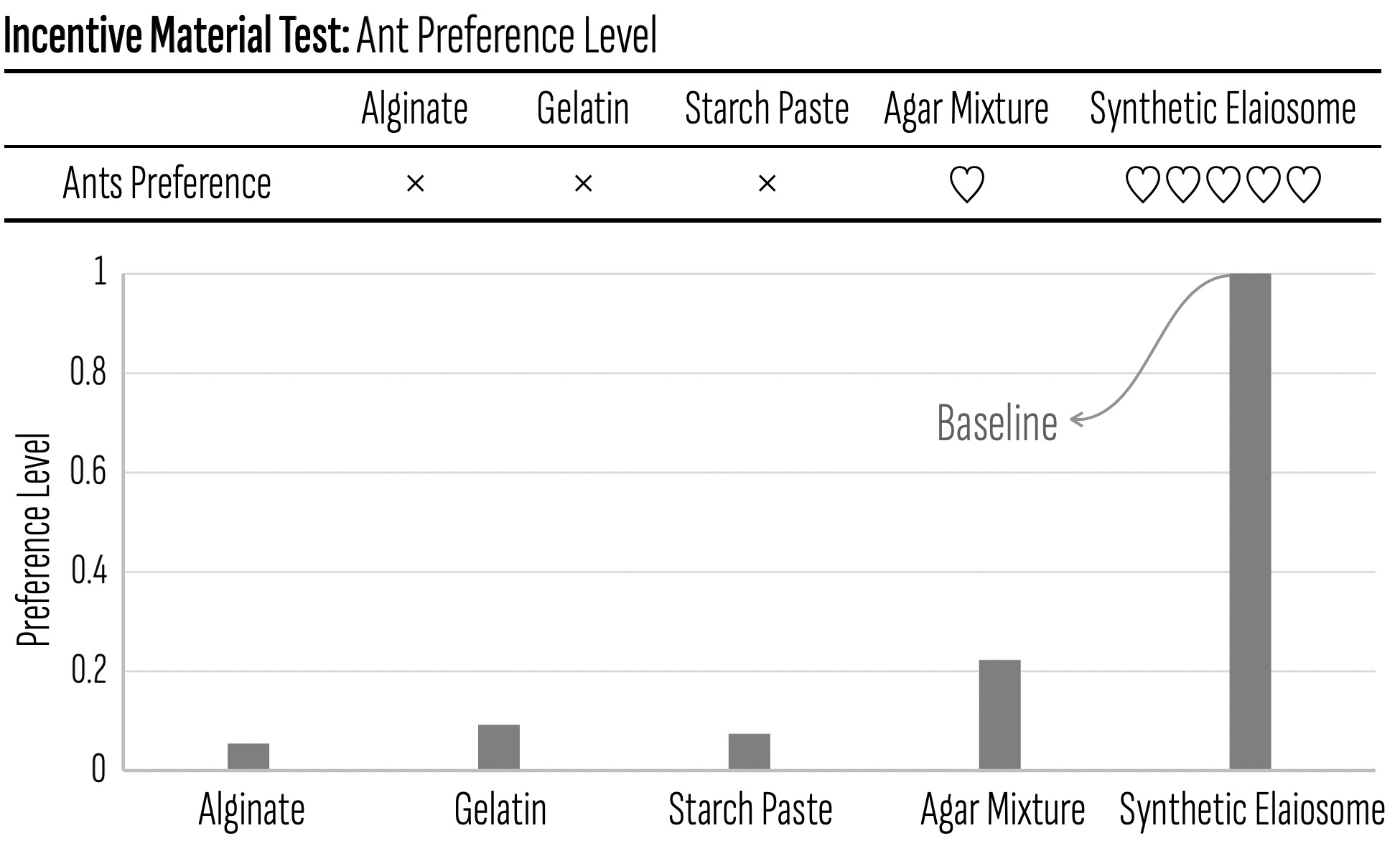}
  \caption{\textcolor{black}{Incentive material test results.}}
  \Description{}
  \label{fig:responsive_material_4}
\end{figure}

\begin{figure*}[t]
  \includegraphics[width=0.95\linewidth]{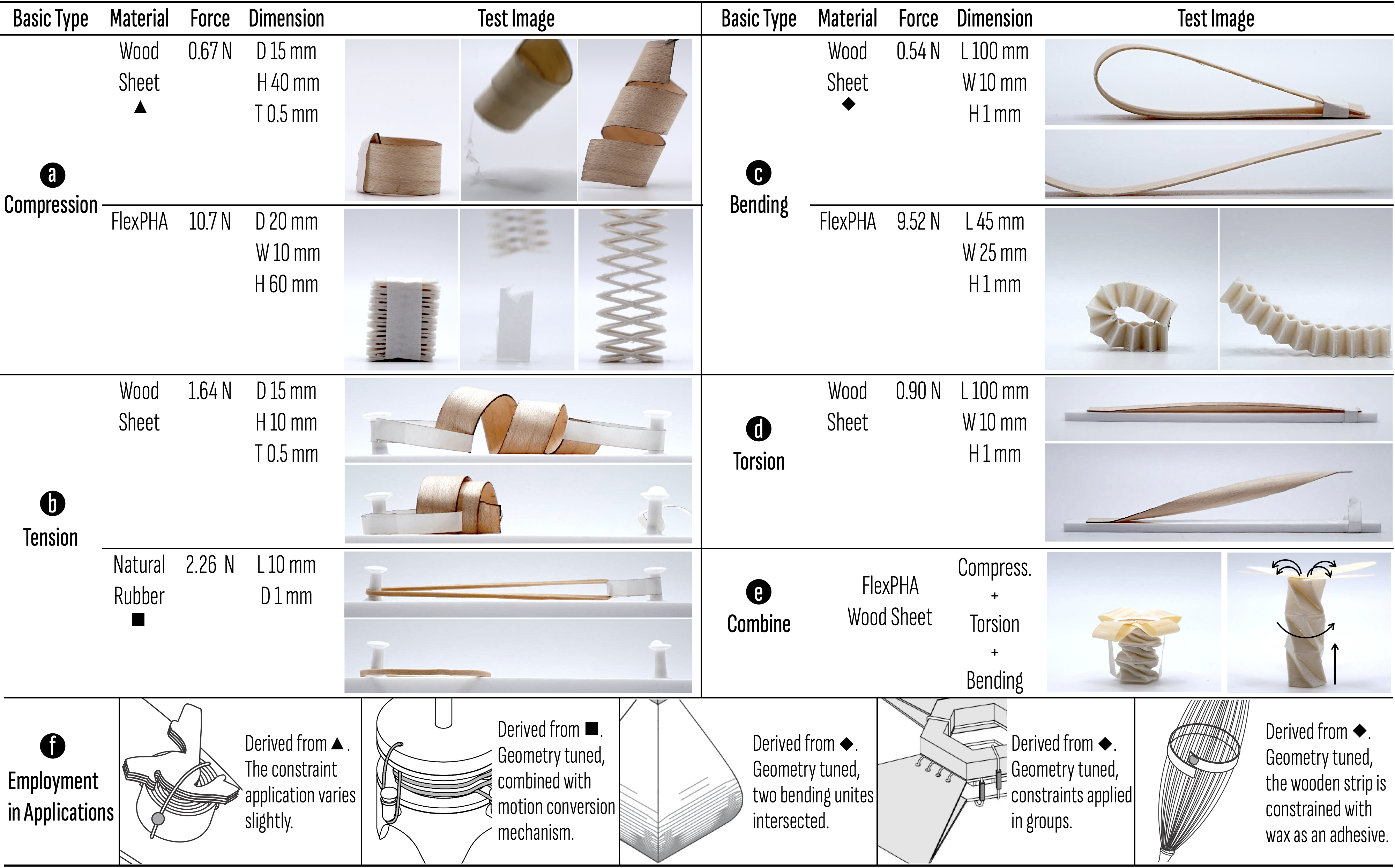}
  \caption{\textcolor{black}{The energy storage (substrate) component and material. Soluble paper serves as the constraint in a to e. The dimensions, including diameter (D), height (H), length (L), width (W), and thickness (T), are measured in a state free of external force. The term 'Force' refers to the stress that the constraint withstands. a-d) Example compression, tension,  bending, and torsion components made of PHA, wood, or natural rubber. e) An example combines compression, bending and torsion. f) Employment of the above components in the application examples.}}
  \Description{}
  \label{fig:substrate_material}
\end{figure*}

\begin{figure*}[b]
  \includegraphics[width=1\linewidth]{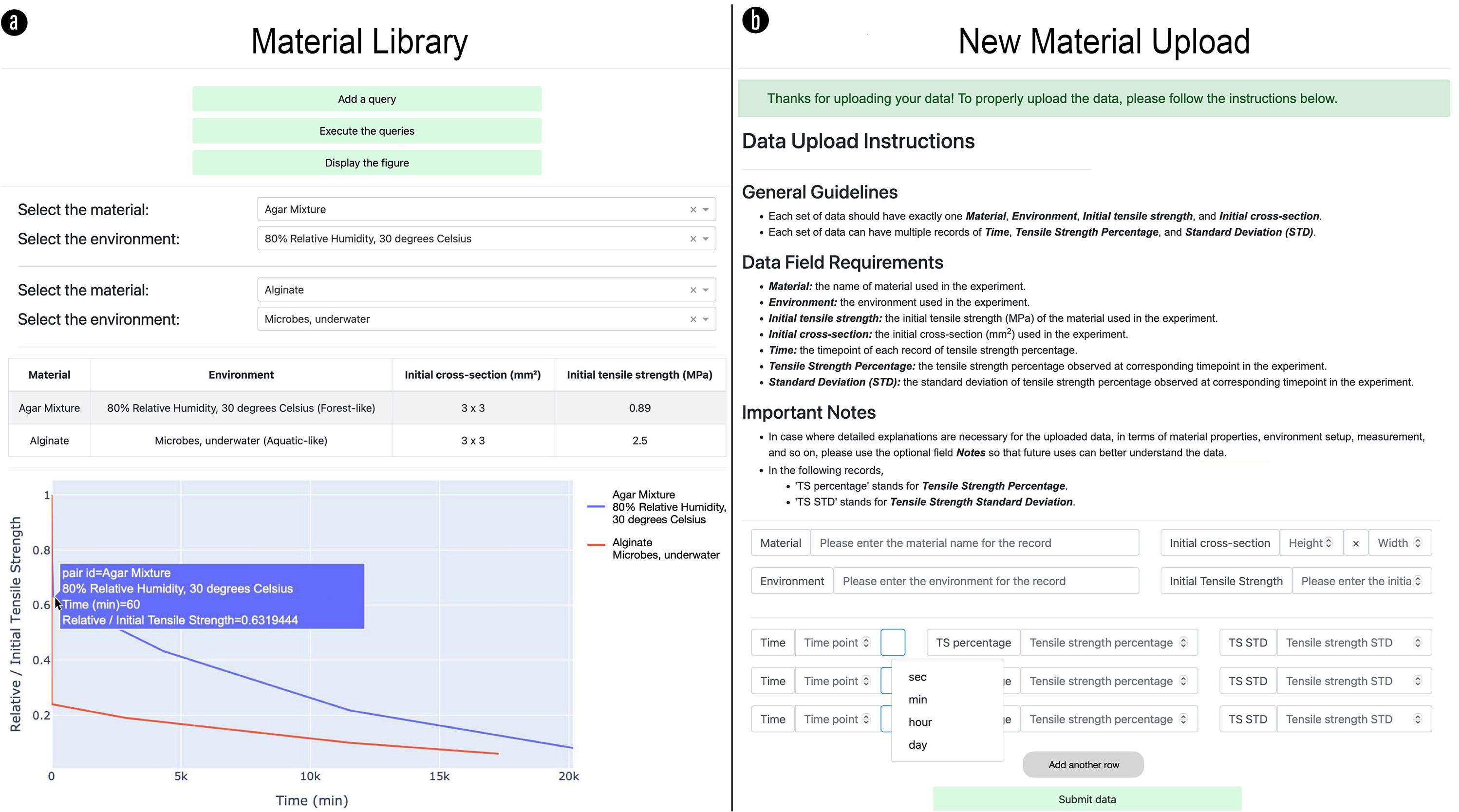}
  \caption{The interactive material library (a) and data input portal (b).}
  \Description{}
  \label{fig:library}
\end{figure*}

\subsubsection{Courier}
We examined the release speed of substances embedded in various materials including alginate, gelatin, CaCO$_3$, and starch pellets. Detailed information regarding the experimental setup is available in Appendix \ref{appendix:B}. The results of the tests conducted on the samples are depicted in Fig. \ref{fig:responsive_material_3}. In general, we observed that starch facilitates the quickest release of content, followed closely by gelatin and CaCO$_3$ which demonstrated comparable release speeds. Alginate exhibited the slowest release speed, with the final 20\% of content being gradually released over a span of first to second days. Furthermore, while the alginate maintained relative stability throughout the three-day testing period, the ones formed by gelatin, CaCO$_3$, and starch pellets experienced a collapse, either dissolving entirely or suspending in the water.

\subsubsection{Incentive}
Building on the composition of elaiosome presented in \cite{konecna_tasty_2018, sasidharan_seed_2019}, we created an elaiosome-like material using a mixture of honey, peanut oil, agar, protein powder, and water in a weight ratio of 2:3:1:1:5. This synthetic elaiosome (s-elaiosome) mimics the fleshy tissue found on many plant seeds, which serves to attract ants. Alongside other edible materials evaluated for use as constraints and couriers, we conducted experiments to assess the attractiveness of our s-elaiosome. As indicated in Fig. \ref{fig:responsive_material_4}, ants showed a markedly higher interest in the s-elaiosome compared to other edible materials. The preparation procedure for the s-elaiosome and the experimental setup are detailed in Appendix \ref{appendix:C}. Notably, the dry s-elaiosome remained stable in room conditions without mold growth, possibly due to natural preservatives in honey and peanut oil \cite{mandal_honey_2011}. Additionally, the material exhibited a tensile strength of $\sim$1 MPa, making it potentially suitable for use as a constraint as well.

\subsection{Substrate Component and Material}
\label{sec:substrate}

For \emph{energy storage} components, \textcolor{black}{Fig. \ref{fig:substrate_material} presents various examples. The force – the stress applied by the constraints to maintain these components in an elastic-energy-stored state – is measured using a HOJILA force gauge (Appendix \ref{appendix:D}). These shapes can serve as a reference in designing such components, with the dimensions shown being adjustable to modify the force; typically, larger dimensions result in greater force.} 

Additionally, we identified PHA (FlexPHA\texttrademark, Beyond Plastic LLC), wood sheets (maple), and natural rubber as suitable materials. Rubber primarily stores elastic energy when stretched, whereas PHA and wood sheets can accommodate various forms of elastic energy storage. PHA is advantageous for 3D printing complex components, while wood sheets are usually laser-cut and can be molded after soaking in water. Wood offers a higher elastic modulus than PHA, allowing it to store more energy under the same geometric conditions. 

\textcolor{black}{PHA is chosen over PLA due to its superior biodegradability in the natural environment \cite{abou-zeid_biodegradation_2004}. PLA typically requires specialized industrial composting conditions to break down efficiently \cite{znaser_polylactic_2021, sudesh_molecular_2000}. However, it's worth noting that substrate materials like PHA, wood, and natural rubber may take months to years to fully degrade, even under "triggered conditions". One may consider collect and reuse components made from these materials until they reach a point where they are no longer functional}.

For \emph{auxiliary} components: PHA, wood and natural rubber can serve as the auxiliary components. Beeswax, carnauba wax and alginate may be used as natural adhesives, depending on the operating temperature or aquatic microbes activity of the device. Natural fibers like cotton thread, feather can further facilitate device assembly as shown in the application examples.

\subsection{Transformation via Degradation}
Transformations mainly take place when the responsive constraint component fails, consequently releasing the energy storage component. The timing of this failure can be predicted using Equ. \eqref{eq:5.3}. Specifically, the constraint component fails when its maximum tensile force $F_{\text{max}}$ falls below the resilience force $F_{\text{res}}$ of the energy storage component due to degradation. The resilience force $F_{\text{res}}$ can be quantified using a force gauge.

Upon the onset of triggered degradation, $F_{\text{max}}$ decreases rapidly. It can be calculated by multiplying the relative tensile strength $\sigma_r$ with the original cross-sectional area $A_0$. Notably, $\sigma_r$ is inversely proportional to time and can be determined experimentally.

\begin{align}
    &\left\{ \begin{array}{ll}
        \text{Compare}(F_{\text{res}}, F_{\text{max}}(t)) \\
        F_{\text{max}} = \sum_{i=1}^{n} (\sigma_r(t) \times A_0) \\
        \sigma_r \propto \frac{1}{t}
    \end{array} \right. \label{eq:5.3}
\end{align}

Given that geometry (e.g., surface-area-to-volume ratio) can affect the rate of material degradation \cite{chamas_degradation_2020, chen_effect_2006}, it's advisable to maintain \( A_0 \) consistent with samples tested in Section \ref{sec:responsive} if one wishes to utilize the test results directly. Then tuning of \( F_{\text{max}} \) can be employed by using multiple (n) constraints components with the same geometry as the tested samples. Similarly, the quantitative results regarding the energy storage component presented in section \ref{sec:substrate} are best suited when a design shares the same geometry and method of constraint application.

However \textcolor{black}{DtF} morphing \textcolor{black}{devices} application scenarios can be very diverse. Sometimes, a customized design with tuned component geometries could be necessary. Fig. \ref{fig:substrate_material}.f shows some structures employed in our application examples, many of which feature tuned components. In such instances, drawing on Section \ref{sec:responsive}, we can still 1) estimate the initial tensile strength of the material to gauge the maximum tension the geometrically-tuned constraint component can provide, 2) identify which material can exhibit more pronounced degradation responsiveness under targeted conditions, and 3) qualitatively estimate the order and time scale (seconds, hours, days) of structural failures. In addition, the various foundational forms shown in Section \ref{sec:substrate} can continue to serve as references for energy storage components. Their geometries can be modified to regulate the force or to add functionalities when needed. Lastly, designs with tuned components could be complemented with experiments to enhance understanding of their quantitative performance, like what we did in Section \ref{sec:application}.

\section{Discussion and Future Work}
\label{Discussion}

\subsection{Extending Trigger Conditions and Material Library}

\textcolor{black}{In addition to the discussed factors, numerous other environmental variables warrant investigation, such as alkalinity and ultraviolet (UV) light. For example, a responsive component might degrade rapidly under high UV levels, triggering the release of a shading structure to protect the device from intense radiation. However, suitable natural materials are often lacking. Materials like alginate \cite{wasikiewicz_degradation_2005}, natural rubber \cite{itoh_effect_2007}, bamboo fiber \cite{rao_photodegradation_2022}, silk fiber \cite{vilaplana_analytical_2015}, and PLA \cite{podzorova_influence_2017} \textcolor{black}{showed no significant tensile strength decline under simulated midday outdoor UV intensity within two weeks \cite{balasaraswathy_uva_2002}.} Although recent studies demonstrate that adding cyclic xanthate to PLA can induce 40\% degradation within six hours of UV exposure \cite{hardy_uv_2022}, its complex chemical synthesis conflicts with our goal. While exploring more environmental factors is promising, progress hinges on advancements in material science, particularly in developing degradable materials through straightforward processes.}

Expanding the material library or even developing a design tool \cite{milliware, MotionFlow, 10.1145/3613905.3648661} is another interesting avenue for research. On one hand, the variety of materials can be increased. For instance, we tested spaghetti and rice noodles; although they are not yet included in our current material selection, their rate of decline in tensile strength in water falls between that of soluble paper and alginate, potentially offering more options for sequence degradation control in water-related applications. On the other hand, fine-tuning existing materials is also possible. For example, our preliminary experiments indicate that adjusting the concentration of sodium alginate or the cross-linking time can impact its degradation rate; alginate with a lower specific surface area often degrades more slowly.

\textcolor{black}{While we've demonstrated how degradation can be cleverly integrated into morphing \textcolor{black}{devices}' functionality, current research on mechanical properties like tensile strength during material degradation is relatively sparse and requires extensive experimental testing.} To address this, we have developed an interactive online database that collates some of our test results, making it easier for users to examine the degradation characteristics of materials under specific conditions (Fig. \ref{fig:library}.a). We have also provided a portal for users to input their own experimental findings into this \textcolor{black}{open-source database}, so we may leverage community input to enrich this database and further broaden the design space for \textcolor{black}{DtF} morphing \textcolor{black}{devices} in the future (Fig. \ref{fig:library}.b). Readers may refer to this \href{https://github.com/qiuyuluuu/DtU}{\textcolor{teal}{link}} for the latest version of the materials library.

\subsection{Performance and Deployment Concerns}
\textcolor{black}{DtF morphing devices} are analog devices based on natural degradation processes, which means their control may not be perfectly time-precise. The deviation in mechanical strength loss can vary from minutes to days, depending on the duration of the triggered degradation. Furthermore, ever-changing real-world environmental conditions can also introduce variability in degradation rates. When designing \textcolor{black}{DtF} applications, it's important to consider whether potential timing inaccuracies could have a significant impact on the intended functionalities.

Additionally, although tests show that the materials used in this work can maintain mechanical properties for at least three months under room conditions, Hartmann et al. \cite{hartmann_becoming_2021} suggest that to guarantee the viability of a device made of biodegradable material, a shelf life of less than one month is not sufficient; less than six months is moderately sufficient, and more than six months is deemed sufficient. Therefore, conducting additional tests and making improvements accordingly to extend the shelf life of these materials can be a good direction for enhancing their practical utility.

Several examples in this paper emphasize enhancing packaging density by controlling the shape of \textcolor{black}{DtF} devices, aiming to improve transportation efficiency and reduce energy consumption and carbon emissions during large-scale deployment. However, for \textcolor{black}{real-world} deployments, it's vital to conduct a careful assessment of their environmental feasibility and impact even though \textcolor{black}{DtF} devices utilize naturally degradable materials: 

1) Strategic Distribution: Deploy a reasonable amount of the devices only in areas and at times where there is a high risk of potential issues, rather than arbitrary distribution. 
 
2)  Retrieval and Reset Decisions: Careful consideration is necessary when deciding whether to retrieve \textcolor{black}{DtF} devices after their functional lifespan.  \textcolor{black}{For some devices, retrieval or reset might be required. For instance, the soil neutralizer may inadvertently release CaCO$_{3}$ over an extended period, as its constraint materials can degrade slowly in non-targeted conditions. In contrast, other devices like the fire monitor may not pose a functional risk if left unretrieved. However, even in cases where retrieval is not functionally necessary, the potential broader impact of introduced materials on the local environment must be considered. Even some natural materials may take months or years to fully degrade, potentially imposing additional burdens on local degradation processes. To address this}, the design stage should prioritize minimizing material usage to limit waste, even if the material is known to be biodegradable locally. Furthermore, the introduction of foreign materials that the local ecosystem may not readily degrade requires careful evaluation to ensure environmental compatibility. 
 
3) Field Testing: Positive in-vitro results are encouraging, but comprehensive field testing is essential to assess their long-term environmental impacts and understand potential side effects before real-world deployment. 
 
 \textcolor{black}{4) Extended Monitoring: Environmental conditions change gradually over time, so it's important to continue monitoring for years after deployment to detect both intended and unintended consequences. This ongoing monitoring helps to understand the effectiveness of the device, allowing for adjustments, and also helps to identify concerning trends early.}

\subsection{Knowledge and Resource Requirement}

\textcolor{black}{Prototyping under the \textcolor{black}{DtF} concept is quite accessible: 1) Equipment and tools: Basic equipment such as low-cost heating plates, laser cutters, or FDM 3D printers suffice for most needs. Where such tools are unavailable, manual crafting with common kitchen utensils is feasible for many designs presented in this paper. Modeling can be accomplished using open-source vector software like Inkscape or free 3D software like Autodesk Fusion. More specialized design and simulation tools could be developed in the future to assist those with less experience. 2) Knowledge: For those unacquainted with material exploration, a basic understanding of material properties and morphing structures would significantly ease the prototyping process. While direct testing of material degradation is practical, leveraging databases, such as the one we’ve shared, opens up possibilities for inclusive user prototyping. Moreover, a deeper understanding of responsive materials can enrich the morphing \textcolor{black}{device} design. For researchers experienced in stimuli-responsive material design and experimentation, the shared \textcolor{black}{DtF} workflows can be valuable in developing their implementation.}

\textcolor{black}{For researchers keen on contributing to the expansion of our database to offer a broader choice of materials, evaluating beyond the metrics discussed in this paper could be beneficial. An interesting aspect to consider is the origin and accessibility of materials. 
\textcolor{black}{Prioritizing materials that are locally abundant and easily obtainable could help reduce carbon footprints and energy demands during material collection and transportation. }
Expanding the scope to encompass the entire lifecycle of materials can lead to more eco-conscious decisions in material selection and application, aligning with broader sustainability goals.} 

\subsection{Designing for Unmaking}
\textcolor{black}{The applications in this paper address the urgent need to engage with ecological transformation and disasters caused by the intersection of climate change and human activities. The DtF strategy enriches the "unmaking" concept by actively engaging in processes of deconstruction and degradation. It expands the idea of degradation from material explorations and structural design to minimize environmental impact, both post-use and during use while providing distinct principles for approaching topics in the early design phase. The DtF strategy aims to reduce dependence on resource-intensive recycling or destruction methods by facilitating a controlled and sustainable unmaking process. This unconventional approach requires a shift in perspective across the entire design process, with the concept of unmaking guiding decision-making at each phase. The DtF morphing device demonstrates a novel approach to HCI aligning with sustainable and ethically conscious principles.}

\section{Conclusion}
In conclusion, the \textcolor{black}{DtF} concept has uncovered a pathway to eco-friendly, self-sustaining morphing \textcolor{black}{devices} operating through environmentally triggered degradations. Through the innovative exploitation of various degradation types and the adept selection of materials and structures responsive to specific environmental triggers, we have fostered a rich design possibility ripe for exploration. Our work exhibits the viability and versatility of the \textcolor{black}{DtF} strategy, showcasing applications across diverse ecosystems and inviting further exploration into this pioneering field. As we forge ahead, we envisage a harmonious blend of technology and environment, unlocking new avenues in sustainable and intelligent design.


\begin{acks}
We extend our gratitude to Lily Yang for her invaluable assistance in maintaining the material library. We also express our appreciation to the reviewers for their constructive feedback, which has significantly contributed to the improvement of this paper. This work was supported by the National Science Foundation under Grants Career IIS2047912 and IIS2017008.

\end{acks}

\balance
\bibliographystyle{ACM-Reference-Format}
\bibliography{reference}

\balance


\begin{figure*}[t]
  \includegraphics[width=\linewidth]{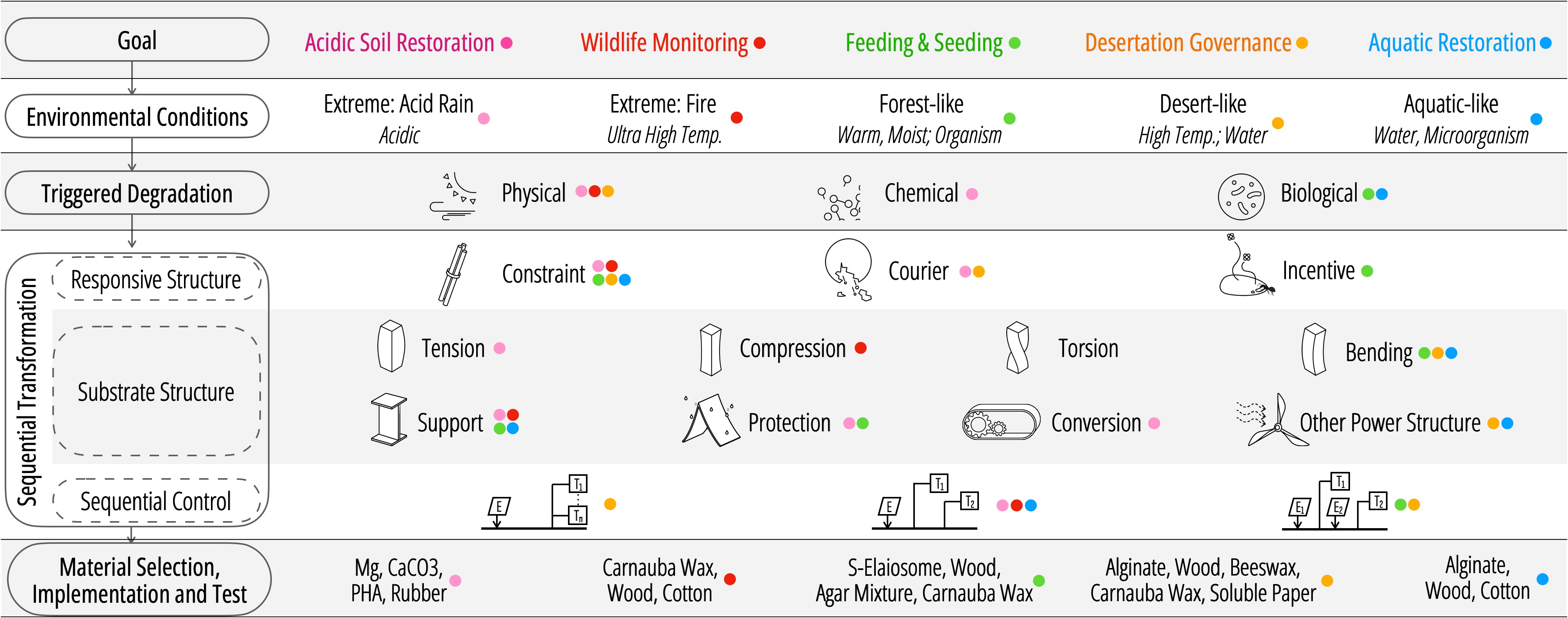}
  \caption{\textcolor{black}{The \textcolor{black}{DtF} design elements and considerations employed by each application example.}}
  \Description{}
  \label{fig:design_flow}
\end{figure*}

\newpage

\appendix
\section{Additional Application Information} \label{appendix:Application}

\subsection{Adaptation of Design Considerations} \label{appendix:design_consideration}
\textcolor{black}{Fig. \ref{fig:design_flow} illustrates
how various design considerations are leveraged in developing the applications examples.}

\subsection{Additional Results}
\textcolor{black}{Here, we present additional experimental results of selective application examples demonstrating their performance in achieving restoration and rehabilitation goals.}

\begin{figure}[b]
  \includegraphics[width=1\columnwidth]{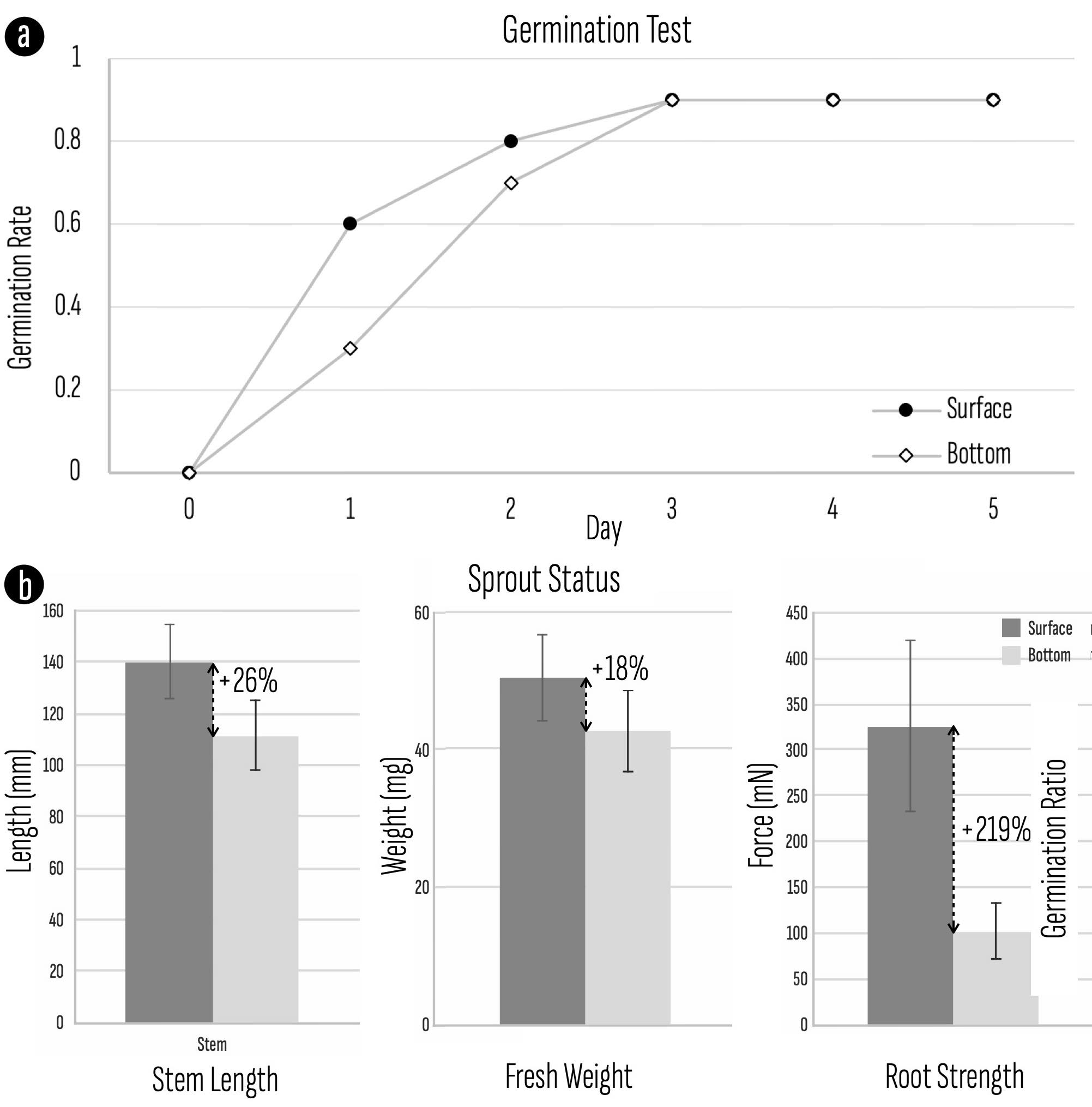}
  \caption{\textcolor{black}{Results from the Germination and Growth Control Experiment: Buoyed seed pods showed enhanced germination rates (a) and superior subsequent plant growth, assessed by the sprouts' stem length, fresh weight, and root strength (b).}}
  \Description{}
  \label{fig:app_ocean_2}
\end{figure}

\textit{Aquatic Habitat Restoration}. Two small aquarium setups were prepared: Group 1 had clay gravel and seeds at the bottom, while Group 2 utilized a seed pod containing both elements. Both groups were filled with an equal amount of water, ensuring that the seeds were submerged and then exposed to natural sunlight. To mimic a coastal water depth of 1 m, Group 1 was partially shaded, reducing sunlight intensity by $\sim$50\% \cite{webb_65_nodate}. After ten days, various growth metrics, including rooting strength, stem length, and fresh weight, were measured. Root strength was evaluated by measuring the maximum force required to pull out the aquatic grass. The results shown in Fig. \ref{fig:app_ocean_2}.a, b indicate that plants in the buoyed seed pods demonstrated superior growth.

\begin{figure}[t]
   \includegraphics[width=1\columnwidth]{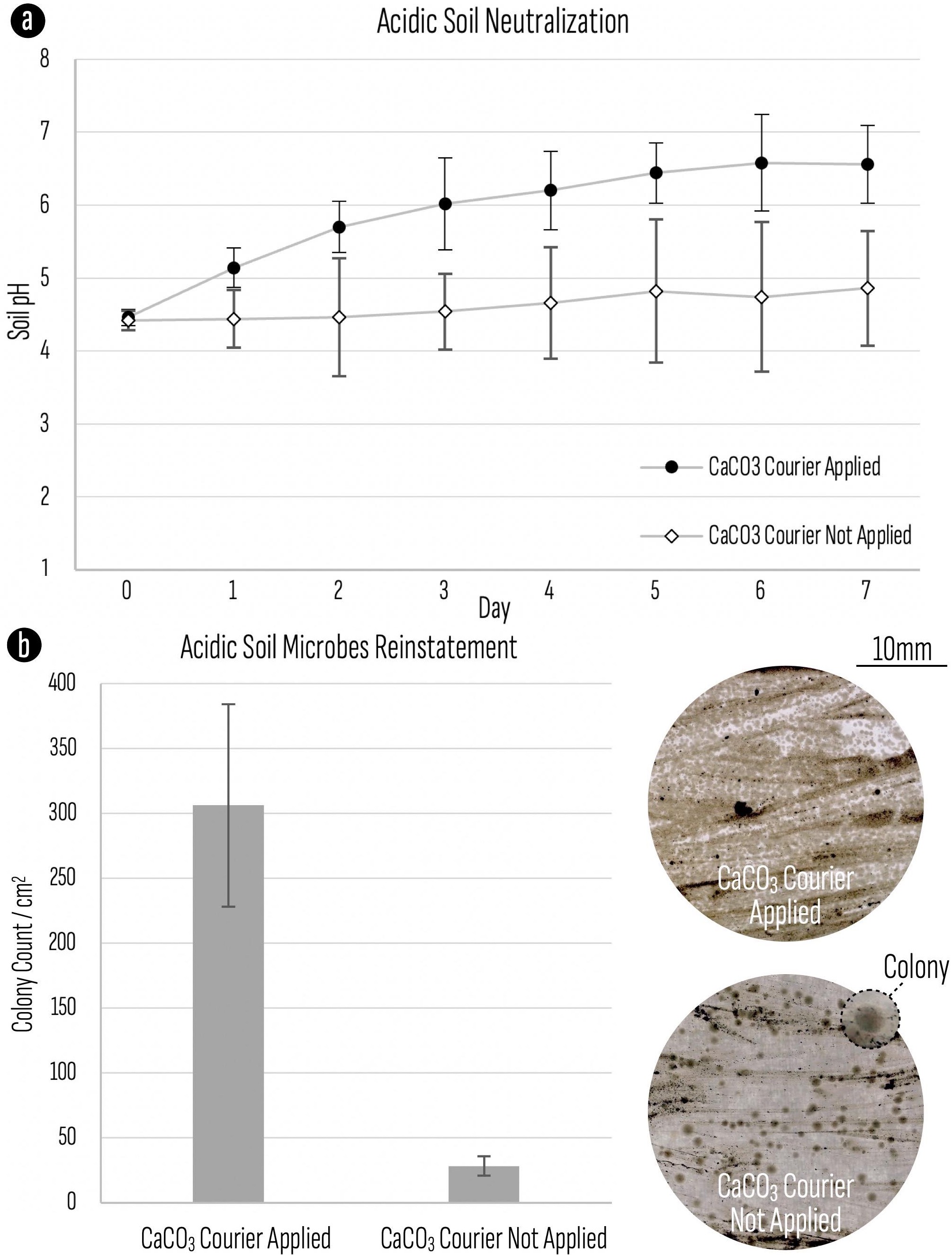}
   \caption{\textcolor{black}{Controlled experiment results for acidic soil restoration: a) soil neutralization outcomes showing the group with CaCO$_3$ couriers achieving faster pH level restoration; b) microflora reinstatement outcomes with significant enhancement in microbial activity observed in the group with CaCO$_3$ couriers. Accompanying photos present a comparison of colony cultures from soil suspensions.}}
   \Description{}
   \label{fig:app_soil_2}
 \end{figure}

\textit{Acidic Soil Monitoring and Rehabilitation}. Healthy soil samples were treated with a pH 4, H$_2$SO$_4$ solution and monitored until the pH stabilized below 5 for three consecutive days. The soil was then split into two portions and placed in dishes with 1.5 cm soil layers. For one portion, CaCO$_3$ couriers (diameter 5 mm, 50 mg Microbes and 50 mg CaCO$_3$) were distributed every \(1 \, \text{cm}^2\). Both samples were watered daily with equal amounts of DI water. pH levels were monitored daily for seven days using a five-point sampling method (5-PT M). On day seven, soil suspensions were swabbed onto petri dishes and colonies in a \(1 \, \text{cm}^2\) area were counted the next day using the 5-PT M. As illustrated in Fig. \ref{fig:app_soil_2}.a, b, the CaCO$_3$ couriers accelerated not only the pH restoration of the acidified soil but also significantly increased microbial activity.

\section{Constraint Material Test Method} \label{appendix:A}
\subsection{Experiment Setup} \label{appendix:A1}
Here, we outline the detailed setup for the constraint material test conducted under different environmental conditions, including one that is not described in section 5.1. The conditions were as follows:

\begin{enumerate}
    \item 40\% \ RH at 25$^\circ$C
    \item 80\% \ RH at 30$^\circ$C
    \item 20\% \ RH at 40$^\circ$C
    \item 0\% \ RH at 70$^\circ$C
    \item Underwater with pH 4 at 25$^\circ$C
    \item Underwater with microbes at 25$^\circ$C
    \item UV at 40\% \ RH at 25$^\circ$C
\end{enumerate}

For conditions (1)-(4), we connected a humidifier to an acrylic container using a hose and affixed a heating pad on top of the container. Both the humidifier and the heating pad were managed by a temperature-humidity controller (Pymeter, PY-20TH) equipped with a sensor placed inside the container to monitor the RH and temperature levels continuously. For condition (7), we employed an additional UVA-340 lamp (Longpro) to simulate the primary UV rays that reach the Earth's surface. The lamp was positioned at a distance that would yield a UVA intensity of $\sim$4 mW/cm$^2$ at the samples. This setup maintained a 12-hour daily light cycle.

For condition (5), we adjusted the pH of 1L of DI water to 4 using 1 mol/L H$_{2}$SO$_{4}$, before transferring it to a 2L tank. In the case of condition (6), 1L of freshwater sourced from a healthy aquarium system was poured into a 2L tank. For both these setups, we utilized a heating pod and a temperature controller (Elitech STC-1000) to maintain the desired temperature. All setups were housed in a room with air conditioning set to 20$^\circ$C and 30\%\ RH.

\subsection{Tensile Strength Testing}

\begin{figure}[h]
  \includegraphics[width=\columnwidth]{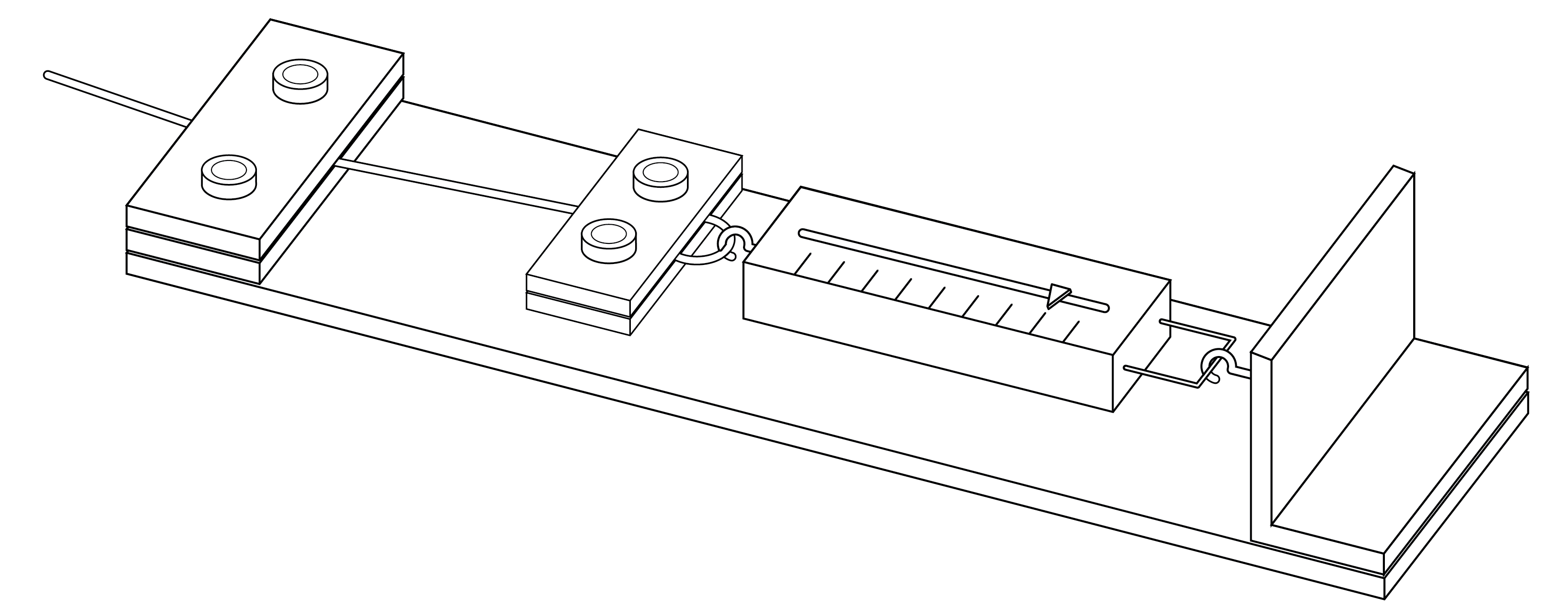}
  \caption{The tensile strength testing platform. The clamps were equipped with soft silicone pads to secure samples with irregular, non-flat shapes. It was vertically mounted to stand upright when in use.}
  \Description{}
  \label{fig:tensile}
\end{figure}

We constructed a test platform depicted fig. \ref{fig:tensile} to facilitate tensile strength testing. A force gauge (HOJILA) integrated into the platform recorded the maximum force exerted from the initiation of load application until the material experienced breakage. Other test procedures are discussed in section 5.1.

\begin{figure}[b]
  \includegraphics[width=\columnwidth]{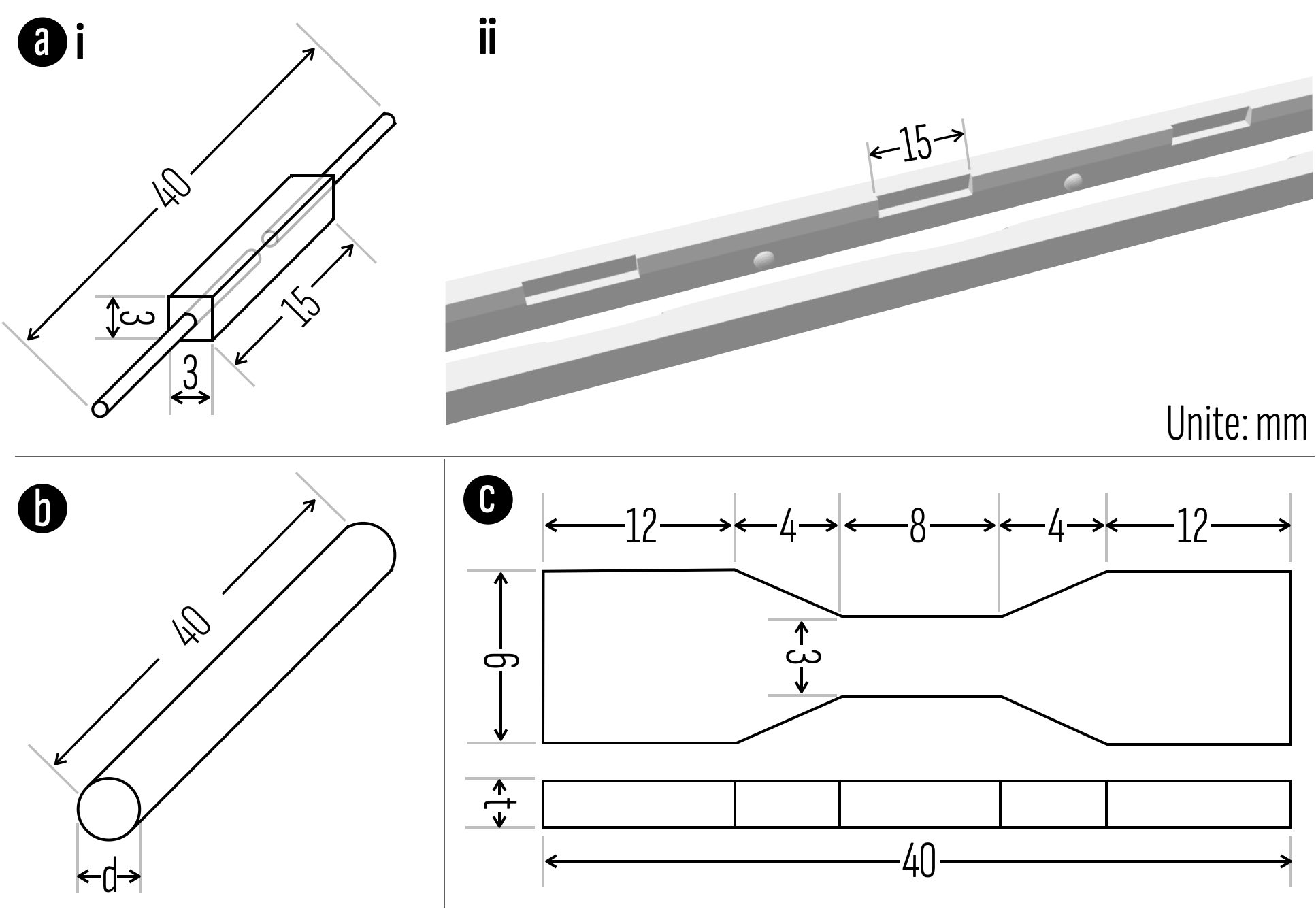}
  \caption{The sample dimension. a) The dimension of casted samples (i) and the mold (ii). b) The dimension of rod samples. c) The dimension of cut samples. t is the thickness of the raw material.}
  \Description{}
  \label{fig:mold}
\end{figure}

\subsection{Sample preparation}\label{appendix:A2}

\textbf{I. General procedure.} For castable materials, as many of them can not stand the testing platform’s clamp, they were prepared and poured into molds that contained pre-placed cut cotton and were left to solidify before being removed from the molds (Fig. \ref{fig:mold}.a). The cast material dimensions were 3 mm x 3 mm x 15 mm with a cotton thread diameter of 1 mm. For materials in the form of thin rods or threads, they were tested in their original shapes. The diameter (d) varied depending on the raw material (Fig. \ref{fig:mold}.b). Other uncastable materials and metals were cut into a dumbbell shape with a middle section width of 3 mm; the thickness (t) varied depending on the material sheet (Fig. \ref{fig:mold}.c).

\begin{figure*}[t]
  \includegraphics[width=1\linewidth]{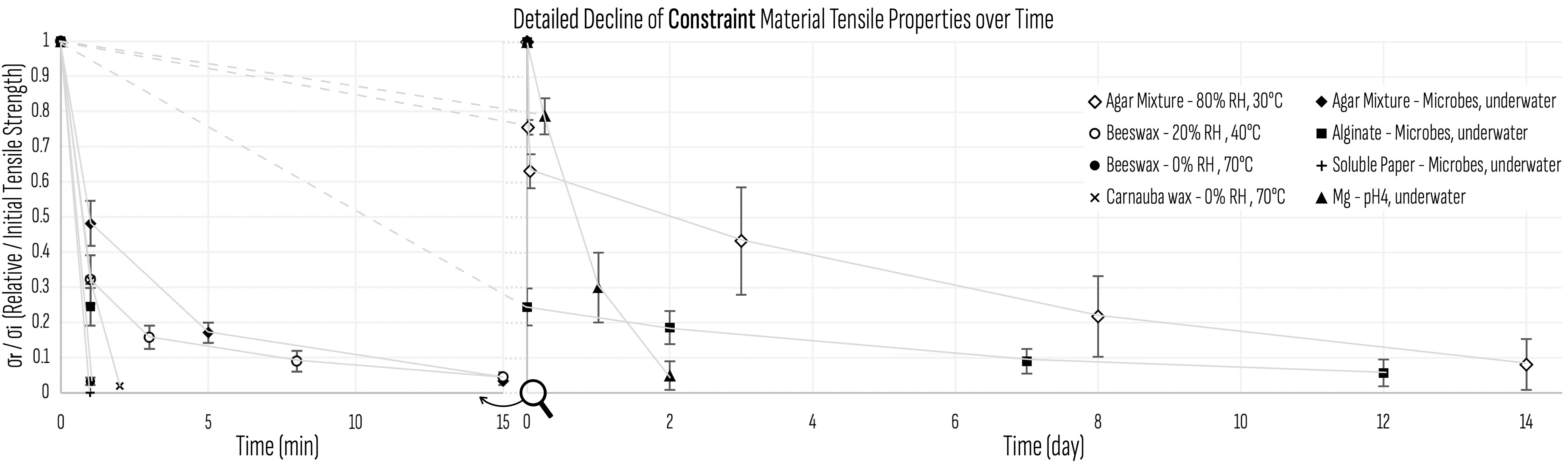}
  \caption{Detailed decline of constraint materials' tensile properties over time.}
  \Description{}
  \label{fig:responsive_material_5}
\end{figure*}

\textbf{II. Detailed procedure.}
\begin{itemize}[leftmargin=10pt]
\item Agar Mixture (2\%\ w/v Sabouraud Dextrose Agar, 15\%\ w/v Gelatin, 5\%\ w/v Dextrose)
     \begin{enumerate}
      \item Soak 15 g of gelatin (PerfectaGel, Gold) in 20 mL of DI water (Aqua Science), stirring continuously for 15 minutes.
      \item Dissolve 2 g of Sabouraud Dextrose Agar (HiMedia, M063-100G) and 5g of dextrose (Nasco) in 80 mL of DI water and bring to a boil to allow complete dissolution.
      \item Remove from heat and cool to below 80$^\circ$C.
      \item Add the gelatin solution to the agar solution and stir for 5 minutes.
      \item Pour into the mold while the solution is above 45°C. Leave at room temperature for 2 hours to solidify.
      \item Remove from the mold and dry with a fan for 48 hours at room temperature.
     \end{enumerate}

\item Alginate
    \begin {enumerate}
    \item Prepare a 3\%\ w/v sodium alginate (Modernist Pantry) solution by dissolving sodium alginate in DI water, stirring at room temperature for 24 hours.
    \item Prepare a 10\%\ w/v calcium chloride (Innovating Science) solution by dissolving it in DI water.
    \item Pour the sodium alginate solution into the mold.
    \item Submerge the mold in the calcium chloride solution for 15 minutes to facilitate alginate crosslinking.
    \item Remove from the mold and dry with a fan for 48 hours at room temperature.
    \end {enumerate}

\item Twizzler
    \begin{enumerate}
    \item Heat Twizzler (Twists Licorice) to 120$^\circ$C until melted.
    \item Spray the mold with cooking oil to minimize cohesion.
    \item Pour the melted Twizzler into the mold without stirring, possibly using a syringe for assistance. Allow it to cool and solidify.
    \item Remove from the mold and dry with a fan for 48 hours at room temperature.
    \end{enumerate}

\item Gummi Candy
    \begin{enumerate}
        \item Heat Gummi Candy (Fruidles) to 160$^\circ$C until it melts.
        \item The follow steps are as same as Twizzler
    \end{enumerate}

\item Rice Paper
    \begin{enumerate}
        \item Soak the rice paper in warm water for 5 minutes.
        \item Cut into a dumbbell shape using a knife.
        \item Allow to dry with a fan at room temperature for 48 hours.
    \end{enumerate}

\item Synthetic Elaiosome
    \begin{enumerate}
        \item Mix 2 g of egg white powder (It's Just!), 6 g of peanut oil (Planters), and 4 g of raw honey. 
        \item The egg white powder can be prepared using a dehydrator and blender. Prepare 10 ml of a 10\%\ w/v Agar solution.
        \item Remove the agar solution from the heat, add the mixture, and stir for about 3 minutes.
        \item Pour into the mold and allow it to set at room temperature for 2 hours.
        \item Remove from the mold and dry with a fan for 48 hours at room temperature.
    \end{enumerate}
    
\item Starch Paste
    \begin{enumerate}
        \item Prepare a 20\%\ w/v starch (Clabber Girl) suspension in cold tap water.
        \item In a separate container, bring an equal volume of tap water to a boil, then reduce the heat to a simmer.
        \item Gradually add the starch slurry to the hot water, stirring constantly.
        \item Cook the mixture over low heat for 10 minutes or until it becomes translucent.
        \item Allow the paste to cool to room temperature.
        \item Pour into the mold, allow it to dry for 3 hours, then demold and dry with a fan at room temperature for 48 hours.
    \end{enumerate}

\item Beeswax
    \begin{enumerate}
        \item Melt beeswax (Cargen) in a water bath at 60$^\circ$C until fully liquefied.
        \item Pour into the mold, then allow it to cool and solidify before demolding.
    \end{enumerate}
    
\item Carnauba wax
    \begin{enumerate}
        \item Melt Carnauba wax (TooGet Technology) in a water bath at 80$^\circ$C until fully liquefied.
        \item Pour into the mold, then allow it to cool and solidify before demolding.
    \end{enumerate}

\item Soluble Paper
    \begin{enumerate}
        \item Cut the soluble paper (SmartSolve, 0.5 mm) into the dumbbell shape using a laser cutter.
    \end{enumerate}

\item Mg (and Al, Zn) Foil
    \begin{enumerate}
        \item Cut the Mg foil (Shandong AME Energy Co., Ltd. 0.05 mm) into the dumbbell shape using a cutting plotter.
    \end{enumerate}

\item Limestone
    \begin{enumerate}
        \item Shape the limestone sample using a knife to obtain the required dimensions for testing.
    \end{enumerate}

\item CaCO3 \& Sand Mixture (HiMedia)
    \begin{enumerate}
        \item Gradually add 15 g CaO to 20 ml water, stirring continuously to form a consistent and lump-free mixture,
        \item Add 10 g sand to the mixture, stirring constantly to maintain a homogenous mixture.
        \item Pour into the mold and compress to firm.
        \item Allow to dry at room temperature for 48 hours, then demold and leave at room condition for an additional 7 days.
    \end{enumerate}

\item CaCO$_{3}$ \& CaSO$_{4}$ Mixture
    \begin{enumerate}
        \item Mix 10 g of CaCO$_{3}$ (PURE ORIGINAL INGREDIENTS) and 10 g of CaSO$_{4}$ (MilliporeSigma) 
        \item Add 6 g of water to the mixture, stirring to form a homogeneous paste.
        \item Pour into the mold and compress to firm.
        \item Allow to dry at room temperature for 48 hours.
        \item Demold and dry with a fan at room temperature for an additional 48 hours.
    \end{enumerate}

\item MgO \& MgCl$_{2}$ Mixture
    \begin{enumerate}
        \item Mix 10 g of MgCl$_{2}$ (Dolotrex) and 10 g of MgO (BULKSUPPLEMENTS) powder.
        \item Add 6 g of water to the mixture, stirring to form a homogeneous paste.
        \item Pour into the mold and compress to firm.
        \item Allow to dry at room temperature for 48 hours, then demold and dry with a fan at room temperature for an additional 48 hours.
    \end{enumerate}

\item Bamboo fiber / Silk fiber 
    \begin{enumerate}
        \item Form into a 1 mm diameter thread.
        \item cut to the desired length.
    \end{enumerate}

\end{itemize}

\subsection{Additional Results Information}\label{appendix:A3}

Fig. \ref{fig:responsive_material_5} details the decline of tensile properties of selected material over time. Additionally, below we delineate the outcomes pertaining to the tested materials that are not currently in use in the \textcolor{black}{DtF} morphing \textcolor{black}{device}.
\\- \emph{Twizzler} and \emph{Gummi Candy} showed rapid degradation under conditions of high humidity, elevated temperatures, and submersion in water. Moreover, they attracted ants. However, the materials raised concerns due to the presence of artificial ingredients that might be unhealthy both for the ants and broader ecosystems. Additionally, they exhibited delayed microbial growth compared to the agar mixture, presumably due to preservatives inhibiting bacterial development. 
\\- \emph{Rice noodle} and \emph{spaghetti} exhibit the potential for biodegradation in warm and moist environments. Moreover, when submerged, they weaken significantly within an hour. This behavior highlights their potential for applications for these conditions.
\\- \emph{Starch paste} exhibits properties similar to those of rice noodles but degrades more rapidly when submerged. However, the extraction process from the mold can be challenging. Optimizations in the molding and demolding processes are necessary to streamline production and enhance practicality.
\\- \emph{Limestone}, along with \emph{mixtures of CaCO$_{3}$ \& sand} (akin to concrete), \emph{CaCO$_{3}$ \& CaSO$_{4}$} (akin to chalk), and \emph{MgO \& MgCl$_{2}$}, were all considered as potential candidate materials expected to degrade quickly in acidic environments. However, we discovered that their reaction rates are considerably slow in mildly acidic conditions. Apart from limestone, these mixtures also undergo substantial weakening in neutral water, which is not desired. Additionally, the mixtures exhibit brittleness, which might further compromise their usability. 
\\- \emph{Al} and \emph{Zn} foils were initially contemplated as prospective materials suitable for acidic environments. Nevertheless, both exhibit a markedly slower reaction rate compared to Mg in mildly acidic conditions. Moreover, Al is not typically regarded as a vital nutrient for living organisms. Compounding this, acidic soil can heighten the solubility of Al ions, potentially harming plant life.
\\- All tested materials exhibited no significant reduction in tensile strength under UV conditions within the test period.

\section{Courier Material Test Method}\label{appendix:B}

\textbf{I. Sample preparation.} All courier test samples are designed with a cylindrical pill shape, possessing a volume of 1 cm³ (diameter: 15 mm, height: ~5.66 mm, when hydrated). Each sample uniformly incorporates 250 mg of glucose (Nasco). To fabricate these, a silicone mold with an inner diameter of 15 mm is utilized.

\begin{itemize}[leftmargin=10pt]
    \item Alginate Courier Sample. Prepare an integrated solution containing 4\%\ w/v alginate and 25\% \ w/v glucose. Cast 1 ml of the prepared solution into the mold. Spray a modest amount of a solution comprising 3\% \ w/v CaCl$_{2}$ and 25\% \ w/v glucose onto it. After a minute, immerse the mold in the 3\% \ w/v CaCl$_{2}$ and 25\% \ w/v glucose solution for 30 minutes. Allow it to dry completely in room condition.
    \item Starch Courier Sample. Introduce 250 mg of glucose into the mold and press it down firmly. Gradually add starch in increments of about 0.2 ml, compacting after each addition until the desired height is achieved. Add 0.2 ml of water and stir thoroughly to obtain a paste, followed by another round of compacting. Allow it to dry entirely in room condition.
    \item CaCO$_{3}$ Courier Sample. Follow the procedure outlined for the starch sample, replacing starch with CaCO$_{3}$ powder.
    \item Gelatin Courier Sample. Prepare an integrated solution containing 10\%\ w/v gelatin and 25\%\ w/v glucose. Cast 1 ml of this solution into the mold. Allow it to solidify and dry in room condition.
\end{itemize}

\textbf{II. Testing procedure.} Prepare three samples for each type of courier, totaling twelve samples. Place each sample in a sealed cup containing 50 ml of DI water maintained at 25$^\circ$C. Utilizing a glucose monitor, record the glucose concentration at the following intervals: 5 mins; every 10 mins from 10-60 mins; hourly from 1-6 hours; and at 12, 24, 48, and 72 hours.

\begin{figure}[b]
  \centering
  \includegraphics[width=\linewidth]{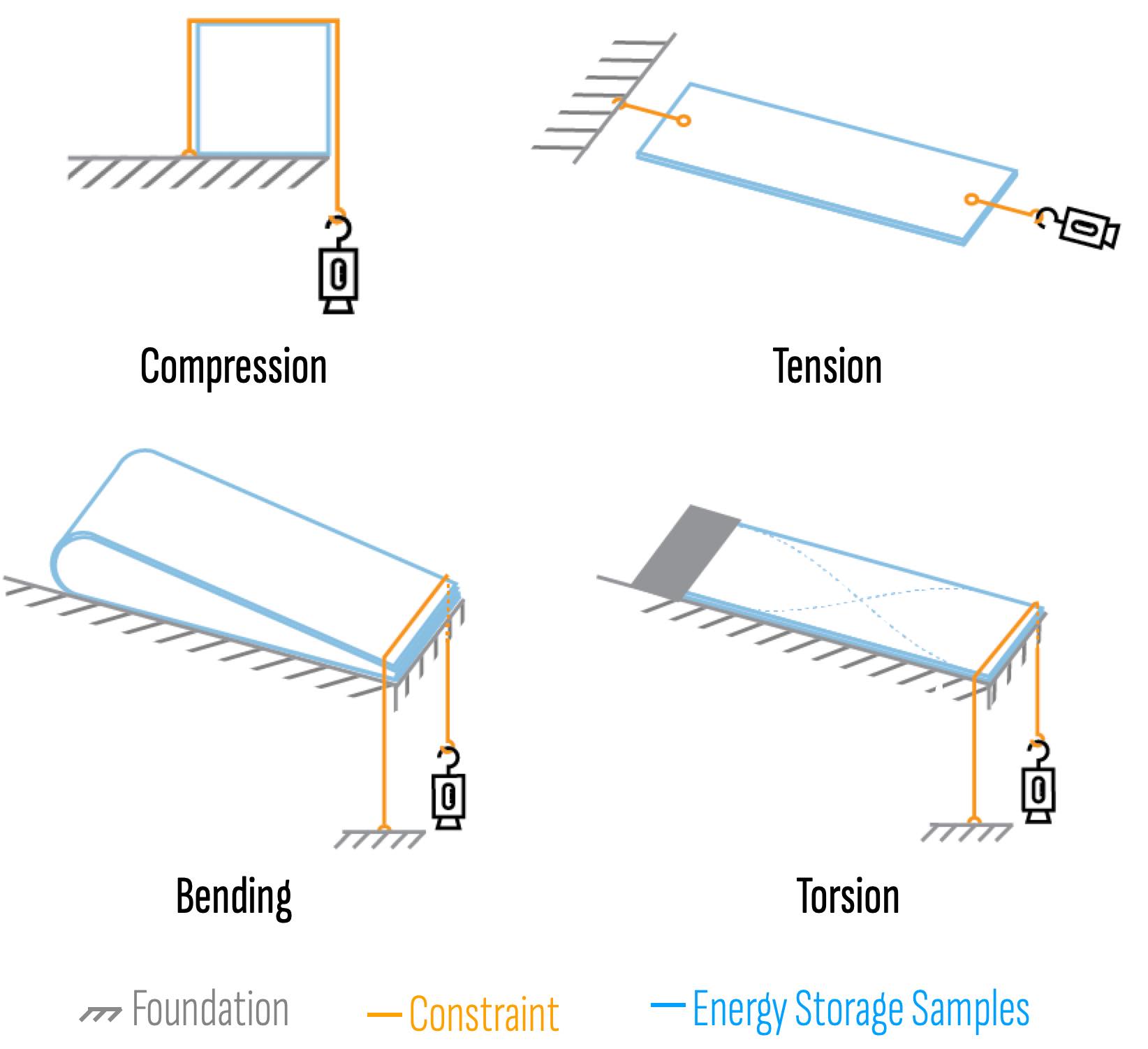}
  \caption{Example setups for measuring the force required to constrain energy storage components.}
  \label{fig:substrate_test}
\end{figure}

\section{Incentive Material Test Method}
\label{appendix:C}

\textbf{I. Sample preparation.} Prepare alginate, gelatin, starch paste, agar mixture, and synthetic elaiosome materials according to the methods previously described in Appendix B1. Pour each mixture into individual molds to form cubes with side lengths of 1 cm each. Allow them to solidify, then dry. 

\textbf{II. Testing procedure.} Identify a suitable ant nest for the experiment. Arrange the five prepared samples equidistantly along the circumference of a circle with a 10 cm diameter, taking the ant nest as the circle's center. Secure each sample to the ground using pins to prevent displacement by ants. Commence a 30-minute observation period immediately after setting up the samples, documenting the number and behavior of ants interacting with each sample throughout this time frame. At the end of the observation period, count the total number of ant visits witnessed for each sample.

\section{Energy Storage Material Test Method}
\label{appendix:D}

\textbf{I. Sample preparation.}
PHA samples can be directly fabricated using 3D printing. For simple structured PHA samples, they can also be 3D printed or cast into PHA sheets first, which can then be cut into desired shapes (such as strips). Subsequently, the required energy storage component shapes can be obtained through heating, molding, and cooling steps. Wood samples are made from commercially available wood veneer. They are laser-cut, then wet, molded, and dried to achieve the desired shape. Natural rubber comes in sheets and is cut to make samples. 

\textbf{II. Testing procedure.} The force required to maintain these components in an elastic energy-stored state is measured using a HOJILA force gauge. Some example setups are illustrated in Fig. \ref{fig:substrate_test}. 

\end{document}